\documentclass[seceq,paper,twocolumn]{jpsj3}

\usepackage{color,bm}
\usepackage{amsmath}
\usepackage{graphicx}

\def\runauthor{
M. Matsumoto, M. Koga, and H. Kusunose
}

\title{
Emergent Odd-Frequency Superconducting Order Parameter \\
near Boundaries in Unconventional Superconductors
}

\author{
Masashige Matsumoto$^1$\thanks{E-mail address: spmmatu@ipc.shizuoka.ac.jp}, Mikito Koga$^2$, and Hiroaki Kusunose$^3$
}

\inst{
$^1$Department of Physics, Faculty of Science, Shizuoka University, Shizuoka 422-8529, Japan \\
$^2$Department of Physics, Faculty of Education, Shizuoka University, Shizuoka 422-8529, Japan \\
$^3$Department of Physics, Ehime University, Matsuyama, Ehime 790-8577, Japan
}

\recdate{October 19, 2012}

\abst{
It was previously suggested that an odd-frequency pair amplitude exists in the vicinity of boundaries in unconventional superconductors.
We develop this idea and quest for a novel superconducting order parameter with an odd-frequency dependence.
For this purpose, we focus on $p$-wave superconductors
and extend the quasi-classical theory to include the odd-frequency dependence in the order parameter.
Both of the frequency and spacial dependences of the order parameter are determined self-consistently.
Under a finite electron-phonon interaction, it is found that an odd-frequency order parameter is stabilized near the boundary
and coexists with the even-frequency one.
By analyzing the induced odd-frequency pair amplitude in terms of the superconducting quasi-particle wavefunction,
it is found that the mid-gap bound state generates the emergent odd-frequency order parameter.
}

\kword{
odd-frequency superconductivity, unconventional superconductivity, chiral $p$-wave superconductivity, spin triplet $s$-wave,
quasi-classical Green's function, Andreev equation, mid-gap state
}

\begin{document}

\maketitle


\newcommand{\Sr}{Sr$_2$RuO$_4$}

\newcommand{\ri}{{\rm i}}

\newcommand{\fx}{{\rm F}_x}
\newcommand{\vf}{v_{{\rm F}}}
\newcommand{\vfx}{v_{{\rm F}x}}
\newcommand{\vfy}{v_{{\rm F}y}}
\newcommand{\vfxa}{v_{{\rm F}x1}}
\newcommand{\vfxb}{v_{{\rm F}x2}}

\newcommand{\rd}{{\rm d}}

\newcommand{\br}{{\bm{r}}}
\newcommand{\bx}{{\bm{x}}}
\newcommand{\bR}{{\bm{R}}}
\newcommand{\brho}{{\bm{\rho}}}
\newcommand{\bG}{{\bm{G}}}
\newcommand{\bDelta}{{\bm{\Delta}}}
\newcommand{\bsigma}{{\bm{\sigma}}}
\newcommand{\bk}{{\bm{k}}}
\newcommand{\bkf}{{\bm{k}_{\rm F}}}
\newcommand{\kfx}{{k_{{\rm F}x}}}
\newcommand{\kfy}{{k_{{\rm F}y}}}
\newcommand{\bd}{\mbox{\boldmath{$d$}}}
\newcommand{\bvf}{{\bm{v_{\rm F}}}}

\newcommand{\tk}{{\theta_k}}
\newcommand{\kf}{k_{\rm F}}
\newcommand{\kfxa}{{k_{{\rm F}x1}}}
\newcommand{\kfxb}{{k_{{\rm F}x2}}}

\newcommand{\om}{{\omega_{m}}}
\newcommand{\ol}{{\omega_{l}}}
\newcommand{\G}{{\hat G}}
\newcommand{\hg}{{\hat g}}
\newcommand{\re}{{\rm e}}

\newcommand{\fbar}{\overline{f}}


\section{Introduction}

Odd-frequency pairing state was suggested by Berezinskii as a possible candidate to describe the superfluid $^3$He.
\cite{Berezinskii}
Although it was not adapted to the superfluid $^3$He,
the idea was developed and applied also to conduction electron systems.
\cite{Kirkpatrick,Belitz-1992,Balatsky,Abrahams-1993,Abrahams-1995,Vojta,Fuseya,Shigeta-2009,Hotta,Kusunose-2011-1,Shigeta-2010,Yanagi}
In these works, the main subject of interest was to understand how the odd-frequency superconductivity is realized in the bulk systems.

From a different point of view, Bergeret {\it et al}. suggested that an odd-frequency pair amplitude is induced
near the interface between superconductor and ferromagnet.
\cite{Bergeret-2001,Bergeret-2003,Bergeret-2005,Eschrig,Yokoyama-2007,Linder-2009,Linder-2010}
Since the magnetic field from the ferromagnet modifies the spin-part wavefunction of the Cooper pair,
it leads to a mixing between the singlet and triplet parings and induces the odd-frequency pair amplitude.
It was also pointed out by Tanaka {\it et al}. that the odd-frequency amplitude is present
also in the vicinity of boundaries in unconventional superconductors.
\cite{Tanaka-2007-1,Tanaka-2007-2,Tanaka-2007-3,Tanaka-2012}.
In this case, the even- and odd-parities of the Cooper pair are mixed by the broken translational symmetry.
This parity mixing is also present around a vortex core and induces the odd-frequency amplitude there.
\cite{Yokoyama-2008,Tanuma}

In the previous works, they focused only on the odd-frequency pair amplitude and did not pay attention to an order parameter.
To understand this, it is important to distinguish between the pair amplitude and the order parameter explicitly.
The pair amplitude ($F$) is defined as an expectation value of the field operators of the Cooper pair
and is understood as a superconducting correlation function,
while the order parameter ($\Delta$) is defined as a product of the pair amplitude and a coupling constant ($V$),
i.e. $\Delta=VF(\Delta)$.
Since the pair amplitude is a function of $\Delta$, the order parameter should be determined in a self-consistent calculation.

In the presence of the pair amplitude, a superconducting order parameter is stabilized
under a finite interaction between conduction electrons.
In fact, we investigated an electron-phonon mediated $s$-wave superconductor
under a finite magnetic field on the basis of the Eliashberg theory
and found that an odd-frequency spin-triplet $s$-wave order parameter
coexists with that of an even-frequency spin-singlet $s$-wave.
\cite{Matsumoto-2012,Kusunose-2012}
The magnetic field leads to the singlet and triplet mixing and induces the odd-frequency pair amplitude.
Since the electron-phonon interaction leads to an attractive interaction also in the spin-triplet $s$-wave channel,
\cite{Kusunose-2011-1}
the odd-frequency order parameter is stabilized.
In other words, it can be understood as a consequence of the even- and odd-frequency mixing of the order parameter
under the broken time-reversal symmetry.
\cite{Matsumoto-2012,Kusunose-2012}
Physical quantities, such as transition temperature, density of states, response to external fields,
show a different behavior owing to the odd-frequency order parameter.
\cite{Matsumoto-2012,Kusunose-2012}

Similarly to the bulk case, we propose in this paper that the even- and odd-frequency order parameters coexist
in the vicinity of boundaries or interfaces of unconventional superconductors.
To demonstrate this, we focus on surface and domain wall in two-dimensional $p$-wave superconductors
and study how the odd-frequency order parameter appears.
To show this, we treat the $p$-wave order parameter on the basis of a weak coupling theory
assuming that the $p$-wave order parameter has no frequency dependence.
On the other hand, we have to take the frequency dependence into account in the odd-frequency order parameter.
It is known that the surface induces an odd-frequency spin-triplet $s$-wave pair amplitude in the $p$-wave superconductors.
\cite{Tanaka-2012}
As the attractive interaction of the spin-triplet $s$-wave, 
we consider an electron-phonon interaction as in our previous study under an external magnetic field.
\cite{Matsumoto-2012,Kusunose-2012}
We determine both the frequency and spatial dependences of the odd-frequency order parameter self-consistently
on the basis of our previous quasi-classical theory for the surface and domain wall in the $p$-wave superconductors.
\cite{Matsumoto-1999}

To understand the emergence of the odd-frequency order parameter,
we express the induced odd-frequency pair amplitude with the superconducting quasi-particle wavefunction.
It enables us to see what type of energy eigenstate generates the odd-frequency pair amplitude.

This paper is organized as follows.
In Sect. 2, the quasi-classical formulation for the odd-frequency order parameter is presented.
The solution of the quasi-classical Green's function is shown in Sect. 3.
We discuss the origin of the emergent odd-frequency order parameter in Sect. 4.
The last section gives summary and discussions.

\section{Formulation}

\subsection{Odd-frequency pair amplitude}

Let us consider a two-dimensional $p_x$-wave superconductor in $x\ge 0$ region
and assume that a specular surface along the $y$-direction is located at $x=0$.
Owing to the translational symmetry along the $y$-direction, the order parameter depends only on $x$.
In this geometry, it is known that there is a mid-gap bound state in the superconducting energy gap near the surface.
The quasi-classical Green's function shows this point clearly.
For a $\bd$-vector parallel to the $z$-axis, the quasi-classical Green's function is separated into two identical $2\times 2$ matrices.
Although the self-consistent $p_x$-wave order parameter has a spatial dependence, we assume a uniform one
to see the induced odd-frequency pair amplitude in a simple way.
For a constant order parameter, the spatial dependent quasi-classical Green's function is expressed as
\cite{Matsumoto-1999,note-d-wave,Matsumoto-1995-2}
\begin{align}
\hg(\bkf,\om,x) = \sum_{i=1,2,3} g_i(\bkf,\om,x)\brho_i,
\label{eqn:g-first}
\end{align}
where $\brho_i$ represents the Pauli matrix of the $i(=1,2,3)$-th component in the Nambu space.
$g_i(\bkf,\om,x)$ is given by
\begin{align}
&g_1(\kfx,\kfy,\om,x) = g_1(-\kfx,\kfy,\om,x) \cr
&~~~~~~~~~~~~~~~~~~~~~~~~
= \frac{| \Delta_x(\bkf) |}{\ri \om} \re^{-2qx}, \cr
&g_2(\kfx,\kfy,\om,x) = - g_2(-\kfx,\kfy,\om,x) \cr
&~~~~~~~~~~~~~~~~~~~~~~~~
                      = - \frac{\Delta_x(\bkf)}{\sqrt{ \om^2 + \Delta_x^2(\bkf) }} \left( 1 - \re^{-2qx} \right), \cr
&g_3(\kfx,\kfy,\om,x) = g_3(-\kfx,\kfy,\om,x) \cr
&~~~~~~~~~~~
                      = \frac{\om}{\sqrt{ \om^2 + \Delta_x^2(\bkf) }}
                        \left[ 1 + \frac{\Delta_x^2(\bkf)}{\om^2} \re^{-2qx} \right],
\label{eqn:g-constant}
\end{align}
with
\begin{align}
q = \sqrt{ \om^2 + \Delta_x^2(\bkf) } / |\vfx|.
\end{align}
Here, $\om=\pi T(2m+1)$ ($m$: integer) is the fermionic Matsubara frequency at temperature $T$.
We take $\hbar=1$ and $k_{\rm B}=1$ throughout this paper.
$\vfx$ represent the $x$ component of the Fermi velocity.
$\bkf=(\kfx,\kfy)$ is the Fermi wave vector and its dependence in the $p_x$-wave order parameter is expressed by $\Delta_x(\bkf)$.
It is characterized by the following symmetry:
$\Delta_x(\kfx,\kfy) = \Delta_x(\kfx,-\kfy) = - \Delta_x(-\kfx,\kfy)$.
We can see that Eq. (\ref{eqn:g-first}) recovers the bulk solution when we take $x\rightarrow\infty$ as
\begin{align}
&\hg(\bkf,\om,{\rm bulk}) \cr
&=
\frac{1}{\sqrt{ \om^2 + \Delta_x^2(\bkf)}}
\left(
  \begin{matrix}
    \om & \ri \Delta_x(\bkf) \cr
    - \ri \Delta_x(\bkf) & \om
  \end{matrix}
\right).
\end{align}
The mid-gap ($E=0$) bound state is described by the second term in $g_3$ in Eq. (\ref{eqn:g-constant}).
It decreases exponentially as $\re^{-2qx}$ for $x\rightarrow \infty$.
Another characteristic point is that $g_2=0$ at $x=0$.
This indicates that the $p_x$-wave order parameter is destroyed by the scattering at the surface and vanishes at $x=0$.

In addition to these, we can see that a finite $g_1$ component appears in the vicinity of the surface.
According to the symmetry with respect to $k_x\leftrightarrow -k_x$ and $k_y\leftrightarrow -k_y$,
the $g_1$ component has an $s$-wave symmetry.
It indicates that the $s$-wave pair amplitude is induced near the surface.
This $s$-wave pair amplitude is a spin-triplet one and has an odd symmetry with respect to $\om\leftrightarrow -\om$.
Thus, the odd-frequency spin-triplet $s$-wave pair amplitude is induced near the surface.
\cite{Tanaka-2012}

\subsection{Odd-frequency order parameter}

As the typical unconventional superconducting state, we focus on two-dimensional $p_x$- and chiral $p_x+\ri p_y$-wave states,
where the latter is suggested in \Sr~superconductor.
\cite{Maeno-1994,Sigrist-2000,Mackenzie,Sigrist-2005,Maeno-2012}
For simplicity, we assume a single cylindrical Fermi surface.
To determine the frequency and spatial dependences of the order parameter,
we use the quasi-classical Green's function introduced by Schopohl {\it et al}. to study the vortex problem.
\cite{Schopohl,Ichioka-1996,Ichioka-1997,Hayashi}
We adapt this scheme to the boundary problem and apply it to both the cases of surface and the domain wall
by extending our previous formulation
\cite{Matsumoto-1999}
to include the odd-frequency order parameter.

The momentum dependent $p$-wave state is given by the $\bd$-vector.
In the bulk region, we assume a case of $\bd(\bk)=(0,0,\Delta_{\rm bulk}(\bk))$
with $\Delta_{\rm bulk}(\bk) = \Delta_x(\bk) + \ri \Delta_y(\bk)$,
where $\Delta_x(\bk)$ and $\Delta_y(\bk)$ are order parameters for the $p_x$- and $p_y$-wave states, respectively.
We take real values for $\Delta_x(\bk)$ and $\Delta_y(\bk)$.
In the bulk region, the Matsubara Green's function is given by the following $4\times 4$ matrix form:
\begin{align}
\bG(\om,\bk) = \left[ \ri \om - \epsilon_\bk \brho_3 - \bDelta_{\rm bulk}(\bk) \right]^{-1}.
\label{eqn:G}
\end{align}
Here, $\bDelta_{\rm bulk}(\bk)$ is the matrix for the order parameter defined as
\begin{align}
\bDelta_{\rm bulk}(\bk) = \Delta_x(\bk) \brho_1 \bsigma_1 - \Delta_y(\bk) \brho_2 \bsigma_1,
\end{align}
where $\bsigma_i$ and $\brho_i$ ($i=1,2,3$) are Pauli matrices for spin and Nambu spaces, respectively.
For the $p_x+\ri p_y$-wave, we apply weak coupling theory assuming that there is no frequency dependence.

Let us discuss an additional order parameter appearing near surface.
In this section, we consider a surface along the $y$-direction.
In this case, the $p_x$-wave order parameter is suppressed near the surface
owing to the odd parity with respect to the reflection at the surface.
Since the spin-triplet $s$-wave pair amplitude is induced,
we introduce the following the additional order parameter for the odd-frequency component:
\begin{align}
\bDelta_{\rm odd}(\bk,\om) = \Delta_s(\bk,\om) \ri \brho_2 \bsigma_1.
\label{eqn:Delta-odd}
\end{align}
Here, $\Delta_s(\bk,\om)$ is a real value.
Owing to the $s$-wave symmetry, we assume that there is no $\bk$ dependence in $\Delta_s(\bk,\om)$.
For the odd-frequency order parameter, it is important to retain the frequency dependence.
The $s$-wave order parameter satisfies
\begin{align}
\bDelta_s(\bk,-\om) = - \bDelta_s(\bk,\om)
\end{align}
reflecting the odd-frequency nature.
The matrix form of the order parameter is then given by
\begin{align}
\bDelta(\bk,\om) &= \bDelta_{\rm bulk}(\bk) + \bDelta_{\rm odd}(\bk,\om) \cr
&=
\left(
\begin{matrix}
  0 & \Delta_1(\bk,\om)\bsigma_1 \cr
  \Delta_2(\bk,\om)\bsigma_1 & 0
\end{matrix}
\right).
\end{align}
There is the following relation between the order parameters:
\begin{align}
\Delta_2(\bk,\om) = - \Delta_1^*(-\bk,\om).
\label{eqn:condition}
\end{align}
This conventional relation holds even when the odd-frequency order parameter is induced by the even-frequency one
and exists as a minority component of the order parameter.
\cite{Matsumoto-2012,Kusunose-2012}
On the other hand, the relation in Eq. (\ref{eqn:condition}) does not hold anymore
if the odd-frequency order parameter is the majority component.
We have to use the other correct form in the latter case.
\cite{Kusunose-2012,note:Berezinskii-rule,Belitz-1999,Solenov,Kusunose-2011-2}
Since the present problem corresponds to the former case, the order parameters are given by
\begin{align}
&\Delta_1(\bk,\om) = \Delta_x(\bk) + \ri \Delta_y(\bk) + \Delta_s(\bk,\om), \cr
&\Delta_2(\bk,\om) = \Delta_x(\bk) - \ri \Delta_y(\bk) - \Delta_s(\bk,\om).
\label{eqn:Delta-12}
\end{align}
There are the following relations:
\begin{align}
&\Delta_1(-\bk,-\om) = - \Delta_1(\bk,\om), \cr
&\Delta_2(-\bk,-\om) = - \Delta_2(\bk,\om), \cr
&\Delta_1(k_x,-k_y,\om) = \Delta_1^*(k_x,k_y,\om), \cr
&\Delta_2(k_x,-k_y,\om) = \Delta_2^*(k_x,k_y,\om).
\label{eqn:symmetry-Delta}
\end{align}

\subsection{Quasi-classical theory with frequency dependent order parameter}

Since the odd-frequency order parameter appears only near the boundary, we have to treat a non-uniform situation.
For this problem, it is convenient to use the quasi-classical Green's function.
For the order parameter given by Eq. (\ref{eqn:Delta-12}), we obtain the following Eilenberger equation:
\begin{align}
&-\ri \bvf \cdot \nabla \hg(\bkf,\om,\br) = \cr
&\left[
  \left(
    \begin{matrix}
      \ri \om & - \Delta_1(\bkf,\om,\br) \cr
      \Delta_2(\bkf,\om,\br) & - \ri \om
    \end{matrix}
  \right),
  \hg(\bkf,\om,\br)
\right], \cr
\label{eqn:green}
\end{align}
where $\hg(\bkf,\om,\br)$ is the quasi-classical Green's function in a $2\times 2$ matrix form.
\cite{Eilenberger,Serene}
The $\bkf$ dependence in $\Delta_1(\bkf,\om,\br)$ and $\Delta_2(\bkf,\om,\br)$
represents the orbital symmetry of the order parameter,
while $\br$ in Eq. (\ref{eqn:green}) represents the center of mass coordinate of the Cooper pair.
The $\om$ dependence in the order parameter is owing to the odd-frequency component.

The quasi-classical Green's function can be written as
\begin{align}
&\hg(\bkf,\om,\br) =
\left(
  \begin{matrix}
    g(\bkf,\om,\br) & \ri f(\bkf,\om,\br) \cr
    -\ri \fbar(\bkf,\om,\br) & - g(\bkf,\om,\br)
  \end{matrix}
\right),
\end{align}
where the components $g(\bkf,\om,\br)$, $f(\bkf,\om,\br)$, and $\fbar(\bkf,\om,\br)$ satisfy the following equations:
\begin{align}
&\bvf \cdot \nabla g(\bkf,\om,\br) = \Delta_2(\bkf,\om,\br) f(\bkf,\om,\br) \cr
&~~~~~~~~~~~~~~~~~~~~~~~~~
                                   - \Delta_1(\bkf,\om,\br) \fbar(\bkf,\om,\br), \cr
& \left( \om + \frac{1}{2} \bvf \cdot \nabla \right) f(\bkf,\om,\br) \cr
&~~~~~~~~~~~~~~~~~~~~~~~
= \Delta_1(\bkf,\om,\br) g(\bkf,\om,\br), \cr
& \left( \om - \frac{1}{2} \bvf \cdot \nabla \right) \fbar(\bkf,\om,\br) \cr
&~~~~~~~~~~~~~~~~~~~~~~~
= \Delta_2(\bkf,\om,\br) g(\bkf,\om,\br).
\end{align}
They can be written as
\cite{Schopohl}
\begin{align}
&g(\bkf,\om,\br) = \frac{ 1 - a(\bkf,\om,\br) b(\bkf,\om,\br) }{ 1 + a(\bkf,\om,\br) b(\bkf,\om,\br) }, \cr
&f(\bkf,\om,\br) = \frac{ 2 a(\bkf,\om,\br) }{ 1 + a(\bkf,\om,\br) b(\bkf,\om,\br) }, \cr
&\fbar(\bkf,\om,\br) = \frac{ 2 b(\bkf,\om,\br) }{ 1 + a(\bkf,\om,\br) b(\bkf,\om,\br) },
\label{eqn:gff}
\end{align}
where $a$ and $b$ satisfy the following equations:
\begin{align}
&\bvf \cdot \nabla a(\bkf,\om,\br) = \Delta_1(\bkf,\om,\br) \cr
&~~~~~~
                                   - \Delta_2(\bkf,\om,\br) a^2(\bkf,\om,\br) - 2 \om a(\bkf,\om,\br), \cr
&\bvf \cdot \nabla b(\bkf,\om,\br) = - \Delta_2(\bkf,\om,\br) \cr
&~~~~~~
                                     + \Delta_1(\bkf,\om,\br) b^2(\bkf,\om,\br) + 2 \om b(\bkf,\om,\br). \cr
\label{eqn:ab}
\end{align}

\begin{figure}[t]
\begin{center}
\includegraphics[width=3.4cm]{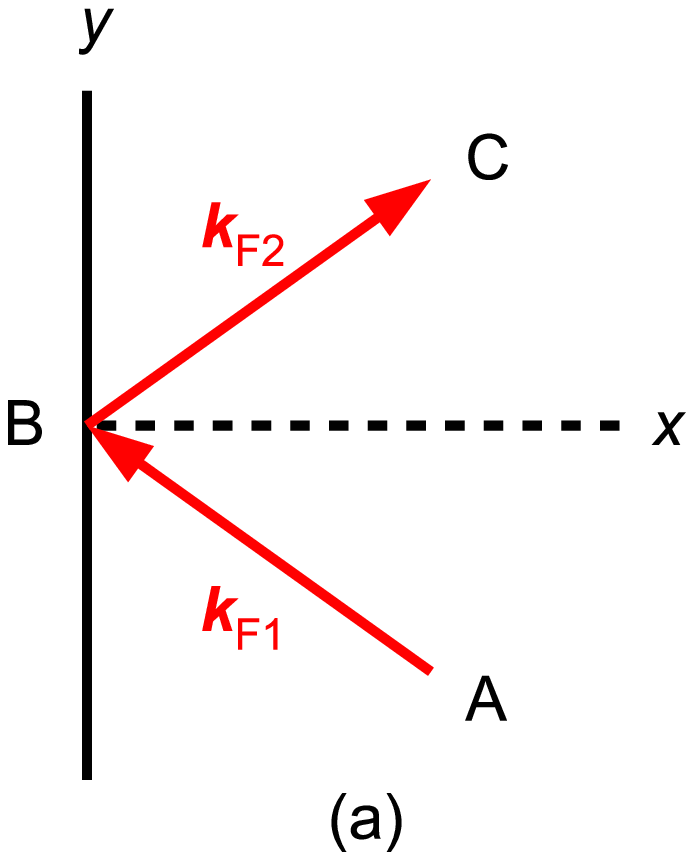}
\hspace{5mm}
\includegraphics[width=4cm]{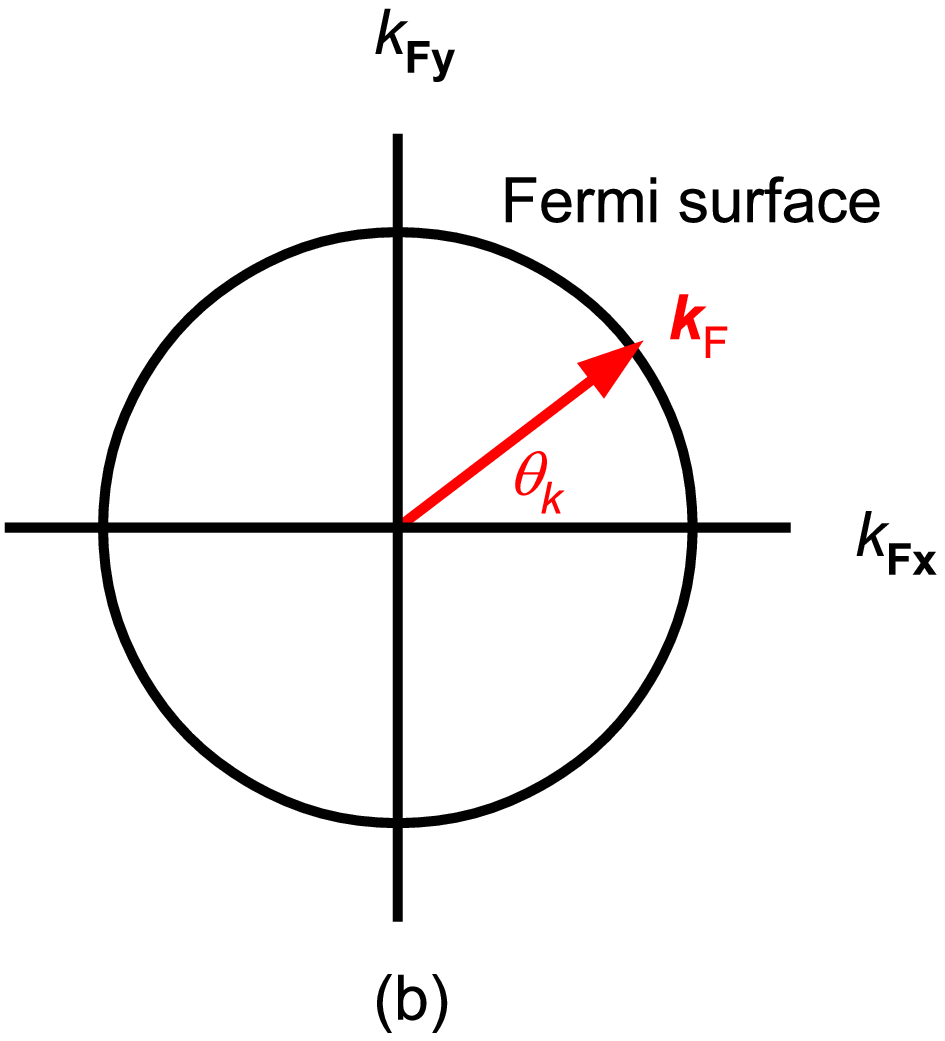}
\end{center}
\caption{
(Color online)
(a) Quasi-classical trajectory of a quasiparticle,
in which the momentum of the incident and reflected quasiparticles along the surface is conserved.
${\bkf}_1$ and ${\bkf}_2$ are the momentum with $\kfx$$<$0 and $\kfx$$>$0, respectively.
A specular surface running along the $y$-direction is located at $x=0$.
(b) Fermi surface and Fermi wavevector for the two-dimensional system.
$\theta_k$ is measured from the $k_x$-axis.
}
\label{fig:surface}
\end{figure}

We solve Eq. (\ref{eqn:ab}) along the quasi-classical trajectory as shown in Fig. \ref{fig:surface}(a),
where the quasiparticle moves from A to B with the momentum $\bk_{\rm F1}$ and from B to C with momentum $\bk_{\rm F2}$.
The Fermi surface on the $k_x-k_y$ plane is cylindrical as shown in Fig. \ref{fig:surface}(b).
Owing to the translational symmetry along the $y$-direction, ${\kfy}$ is conserved.
Then, we match the two solutions using the following boundary condition at point B:
\cite{Serene,Nagai,Hara-1988,Ashida-1989,Nagato-1993}
\begin{align}
\hg(\bkf_1,\om,{\rm B})=\hg(\bkf_2,\om,{\rm B}),
\label{eqn:boundary-G}
\end{align}
which means for $a$ and $b$ as
\begin{align}
&a(\bkf_1,\om,{\rm B}) = a(\bkf_2,\om,{\rm B}), \cr
&b(\bkf_1,\om,{\rm B}) = b(\bkf_2,\om,{\rm B}).
\end{align}
Since the system is translationally invariant along the $y$-direction,
the quasi-classical Green's function or $a(\bkf,\om,\br)$ and $b(\bkf,\om,\br)$ depend only on $x$.
Then, Eq. (\ref{eqn:ab}) can be rewritten as
\begin{align}
&\vfx\frac{\rd}{\rd x} a(\bkf,\om,x) = \Delta_1(\bkf,\om,x) \cr
&~~~~~~
                                     - \Delta_2(\bkf,\om,x) a^2(\bkf,\om,x) - 2 \om a(\bkf,\om,x), \cr
&\vfx\frac{\rd}{\rd x} b(\bkf,\om,x) = - \Delta_2(\bkf,\om,x) \cr
&~~~~~~
                                     + \Delta_1(\bkf,\om,x) b^2(\bkf,\om,x) + 2 \om b(\bkf,\om,x). \cr
\label{eqn:ab-2}
\end{align}
where $\vfx$ is the $x$ component of the Fermi velocity.
The initial and boundary conditions for Eq. (\ref{eqn:ab-2}) are given by
\begin{align}
&a(\bkf_1,\om,x=\infty) \cr
&= \frac{ \Delta_1(\bkf_1,\om,\infty) }{ \sqrt{ \om^2 + \Delta_1(\bkf_1,\om,\infty) \Delta_2(\bkf_1,\om,\infty) } + \om }, \cr
&b(\bkf_2,\om,x=\infty) \cr
&= \frac{ \Delta_2(\bkf_2,\om,\infty) }{ \sqrt{ \om^2 + \Delta_1(\bkf_2,\om,\infty) \Delta_2(\bkf_2,\om,\infty) } + \om }, \cr
&a(\bkf_1,\om,x=0) = a(\bkf_2,\om,x=0), \cr
&b(\bkf_1,\om,x=0) = b(\bkf_2,\om,x=0).
\label{eqn:ab-initial}
\end{align}

We write the superconducting order parameters as
\begin{align}
\Delta_1(\bkf,\om,x) &= \Delta_x(x) \cos\tk + \ri \Delta_y(x) \sin\tk + \Delta_s(\om,x), \cr
\Delta_2(\bkf,\om,x) &= \Delta_x(x) \cos\tk - \ri \Delta_y(x) \sin\tk - \Delta_s(\om,x),
\end{align}
where $\Delta_x(x)$, $\Delta_y(x)$, and $\Delta_s(\om,x)$
are the superconducting order parameters for the $p_x$-, $p_y$-, and $s$-waves, respectively.
The Fermi wavevector dependences are expressed by the angle of $\tk$ as shown in Fig. \ref{fig:surface}(b).
The order parameters are real numbers and are determined by
\cite{Bruder}
\begin{align}
&\left(
  \begin{matrix}
    \Delta_x(x) \cr
    \Delta_y(x) \cr
    \Delta_s(\ol,x) \cr
  \end{matrix}
\right)
= 2 T \sum_{0 < \om < \omega_{\rm c}} \int_0^{\frac{\pi}{2}} \rd \tk \cr
&\times
\left(
  \begin{matrix}
    2 V_x \cos \tk {\rm Re} \left[ f(\tk,\om,x) - f(\pi-\tk,\om,x) \right] \cr
    2 V_y \sin \tk {\rm Im} \left[ f(\tk,\om,x) + f(\pi-\tk,\om,x) \right] \cr
    V_s V_-(\ol,\om) {\rm Re} \left[ f(\tk,\om,x) + f(\pi-\tk,\om,x) \right]
  \end{matrix}
\right), \cr
\label{eqn:gap-equation}
\end{align}
where the following relations were used to derive the gap equation:
\begin{align}
&f(\tk+\pi,-\om,x) = - f(\tk,\om,x), \cr
&f(-\tk,\om,x) = f^*(\tk,\om,x).
\end{align}
In Eq. (\ref{eqn:gap-equation}), $V_x$, $V_y$, and $V_s$ are dimensionless coupling constants
for the $p_x$-, $p_y$- and $s$-waves, respectively.
In the $p_x+\ri p_y$-wave pairing, the coupling constants are expressed as
\cite{Kieselmann}
\begin{align}
V_x = V_y = \left( \log{\frac{T}{T_{\rm c}}} + \sum_{0 < m < \frac{\omega_{\rm c}}{2\pi T}} \frac{1}{m - \frac{1}{2}} \right)^{-1},
\end{align}
where $T_{\rm c}$ is the superconducting transition temperature for the $p$-wave.

Next, we consider an effective interaction for the $s$-wave.
Since the $s$-wave order parameter depends on the frequency, we need a frequency dependence in the interaction.
As the simplest model, we consider Einstein phonons coupled to the conduction electrons.
Since the electron-phonon interaction does not depend on spins of the conduction electrons,
it is attractive for both the singlet and triplet channels.
\cite{Kusunose-2011-1}
The odd-frequency dependence of the effective interaction for the $s$-wave is given by
\cite{Kusunose-2011-1,Matsumoto-2012}
\begin{align}
V_-(\ol,\om) = \frac{\omega_{\rm E}^2}{ ( \ol - \om )^2 + \omega_{\rm E}^2 }
             - \frac{\omega_{\rm E}^2}{ ( \ol + \om )^2 + \omega_{\rm E}^2 },
\label{eqn:V-odd}
\end{align}
where $\omega_{\rm E}$ is the frequency of the Einstein phonon.
We note that $V_-(\ol,\om)$ is antisymmetric with respect to $\ol \rightarrow -\ol$ and $\om \rightarrow -\om$,
i.e. $V_-(-\ol,\om) = V_-(\ol,-\om) = - V_-(\ol,\om)$.
In Eq. (\ref{eqn:gap-equation}), we introduced a cutoff energy $\omega_{\rm c}$ in the Matsubara frequency summation.
In general cases, the cutoff energies are different for the $p$- and $s$-waves,
however, the cutoff energies are irrelevant to discuss the odd-frequency order parameter.
We assume the same cutoff energies for both the $p$- and $s$-wave order parameters here.

Now we can determine the order parameters
by solving Eq. (\ref{eqn:gap-equation}) with Eqs. (\ref{eqn:gff}), (\ref{eqn:ab-2}), and (\ref{eqn:ab-initial}) self-consistently.

\section{Self-Consistent Solution}

\subsection{$p_x$-wave case}

\begin{figure}[t]
\begin{center}
\includegraphics[width=6cm,clip]{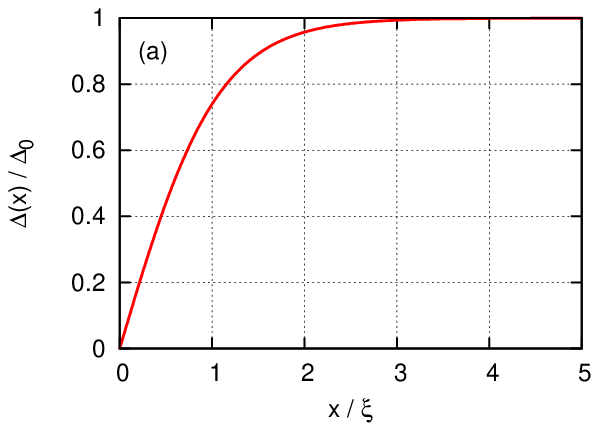}
\includegraphics[width=6cm,clip]{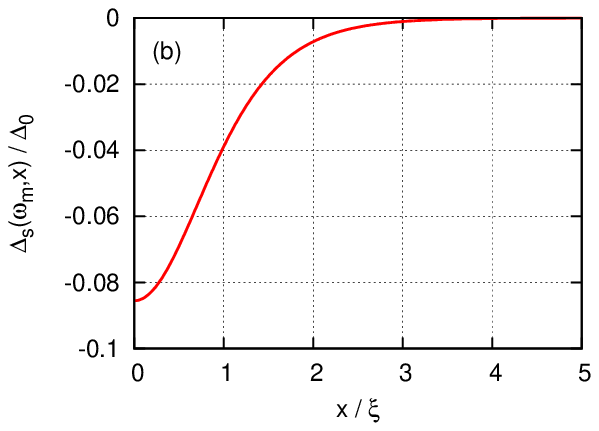}
\includegraphics[width=6cm,clip]{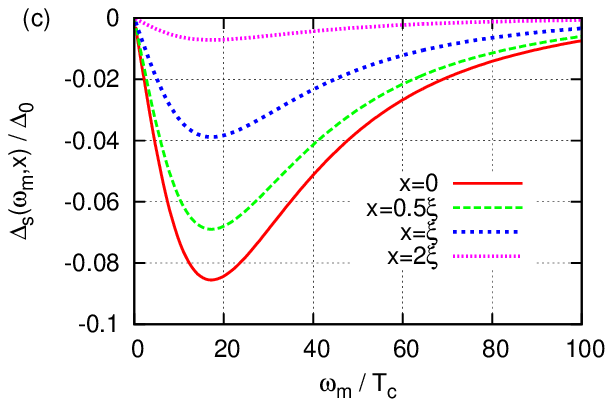}
\includegraphics[width=6.5cm,clip]{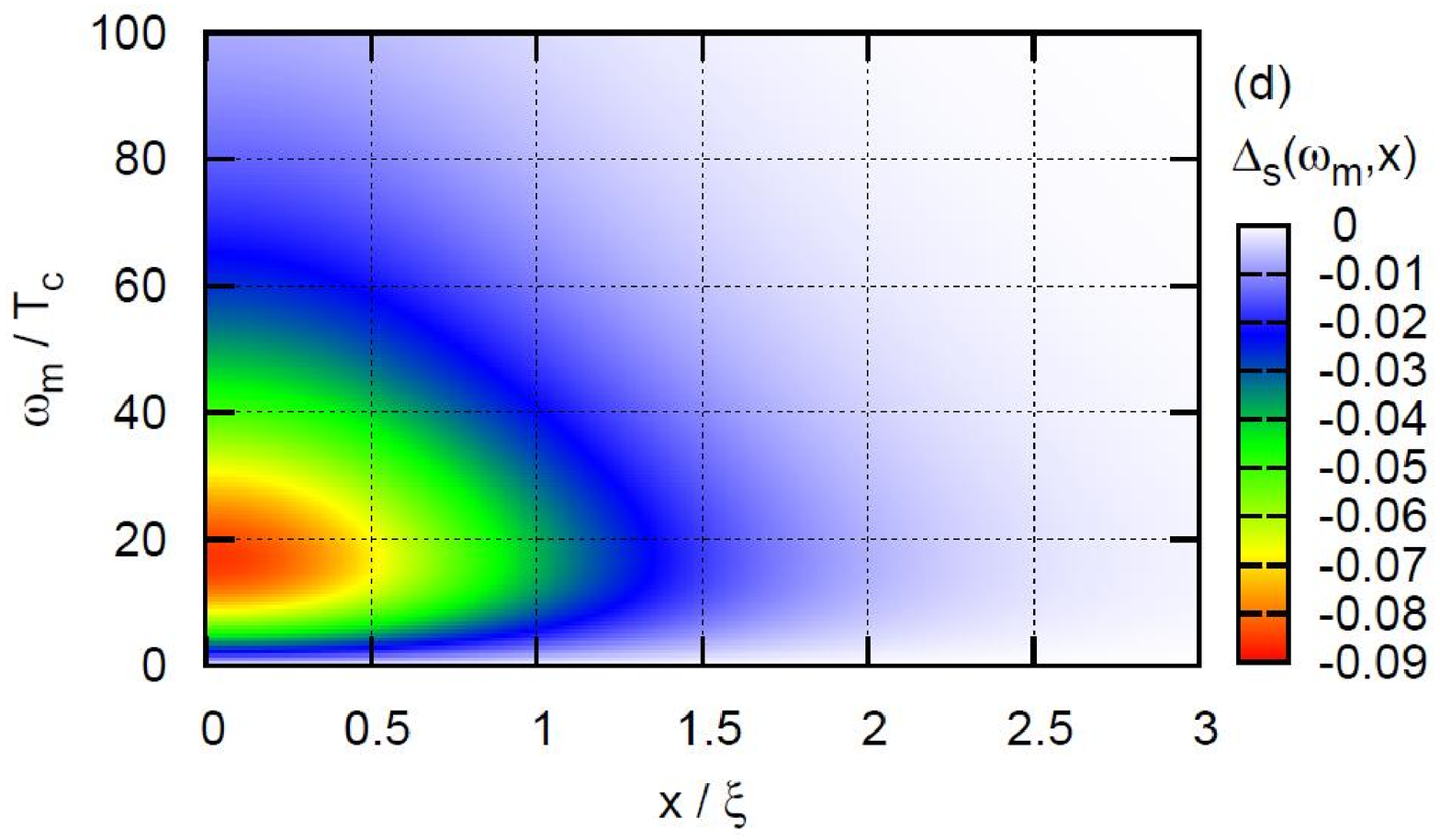}
\end{center}
\caption{
(Color online)
(a) Spatial dependence of the $p_x$-wave order parameter at $T=0.1 T_{\rm c}$ for $V_y=0$ (pure $p_x$-wave).
The order parameters are normalized by the balk value at $T=0$, $\Delta_0=\Delta_x(\infty)\simeq 2.14 T_{\rm c}$.
The $x$ coordinate is measured in unit of the coherence length defined by $\xi = v_{\rm F}/\Delta_0$.
The parameter is chosen as $\omega_{\rm c}=100 T_{\rm c}$.
(b) Spatial dependence of the $s$-wave order parameter.
The Matsubara frequency is fixed as $\om=17.3 T_{\rm c}$ at which $|\Delta_s(\om,x)|$ has the maximum value.
The set of parameters are chosen as $\omega_{\rm E}=10 T_{\rm c}$ and $V_s=10 V_x$.
(c) Matsubara frequency dependence of the odd-frequency spin-triplet $s$-wave order parameter at various positions of $x$.
(d) Contour plot of $\Delta_s(\om,x)$.
}
\label{fig:gap-px}
\end{figure}

First, we study a case of $p_x$-wave assuming $V_y=0$ [$\Delta_y(x)=0$].
Before going the non-uniform solution, let us summarize the bulk solution.
The transition temperature $T_{\rm c}$ and the solution of the gap equation [Eq. (\ref{eqn:gap-equation})]
at $T=0$ for $x\rightarrow\infty$ are obtained as
\cite{Matsumoto-1999,Matsumoto-1995-2}
\begin{align}
&T_{\rm c} = \frac{ 2 \re^\gamma \omega_{\rm c} }{\pi} \re^{-\frac{1}{V_x}}, \cr
&\Delta_x(\infty) = 4 \omega_{\rm c} \re^{-\left( \frac{1}{V_x} + \frac{1}{2} \right) } \simeq 2.14 T_{\rm c} \equiv \Delta_0.
\end{align}
Here, $\gamma$ is the Euler's constant: $\gamma$=0.57721$\cdots$.
For the characteristic length, we define the following coherence length for the superconducting state:
\begin{align}
\xi = \frac{v_{\rm F}}{\Delta_0}.
\end{align}

In Fig. \ref{fig:gap-px}, we show the self-consistently determined order parameters.
The order parameters and the spatial coordinate are scaled by $\Delta_0$ and $\xi$, respectively.
The set of parameters are chosen as $\omega_{\rm c}=100 T_{\rm c}$, $\omega_{\rm E}=10 T_{\rm c}$, and $V_s=10 V_x$.
We can see in Fig. \ref{fig:gap-px}(a) that $\Delta_x(x)$ decreases near the surface and disappears completely at $x=0$.
\cite{Hara-1986}
In contrast to this component, the odd-frequency $s$-wave order parameter $\Delta_s(\om,x)$ appears in the vicinity of the surface
as shown in Fig. \ref{fig:gap-px}(b).
Its magnitude has a maximum value at $x=0$ and decreases with $x$ monotonically.
In Fig. \ref{fig:gap-px}(c), we show the Matsubara frequency dependence of the $s$-wave order parameter at various positions of $x$.
Since the $s$-wave order parameter must have an odd-frequency dependence owing to the fermion property,
$\Delta_s(\om,x)$ shows a linear $\om$ dependence for small $\om$
and has a peak around $\om \sim \omega_{\rm E} = 10 T_{\rm c}$ as expected.
\cite{Kusunose-2011-1,Matsumoto-2012,Kusunose-2012}
In Fig. \ref{fig:gap-px}(d), we show the contour plot of $\Delta_s(\om,x)$.
When we compare Figs. \ref{fig:gap-px}(a) and \ref{fig:gap-px}(b),
we notice that the magnitude of the odd-frequency order parameter is quite small
even under a strong attractive interaction ($V_s=10 V_x$).
Therefore, the feedback effect of $\Delta_s(\om,x)$ to $\Delta_x(x)$ is very weak
and that the value of $\Delta_x(x)$ is almost unchanged from that in the absence of $\Delta_s(\om,x)$ ($V_s=0$ case).

\begin{figure}[t]
\begin{center}
\includegraphics[width=6.5cm,clip]{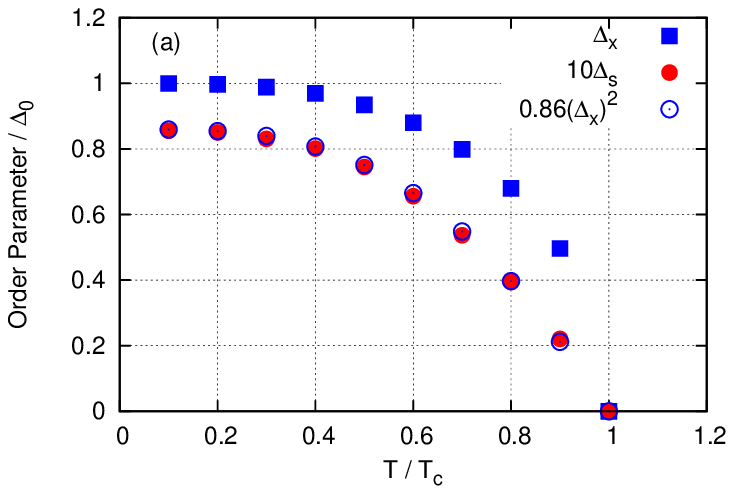}
\includegraphics[width=6.5cm,clip]{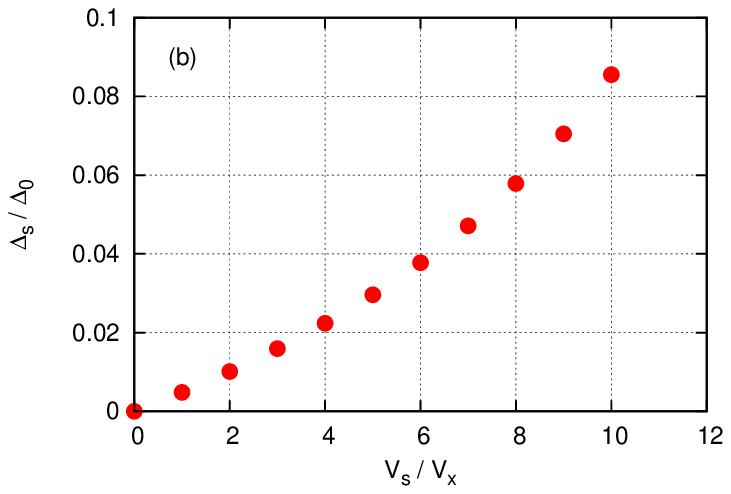}
\end{center}
\caption{
(Color online)
(a) Temperature dependence of the order parameters scaled by $\Delta_0$.
$\Delta_x$ is the $p_x$-wave component in the bulk region ($x\rightarrow \infty$).
$\Delta_s$ is the maximum magnitude of $|\Delta_s(\om,x)|$ at $x=0$, where the maximum value appears at $\om\sim 17 T_{\rm c}$.
(b) Coupling constant ($V_s$) dependence of the maximum value of $|\Delta_s(\om,x)|$ at $x=0$ for $T=0.1 T_{\rm c}$.
}
\label{fig:gap-temp-px}
\end{figure}

We show the temperature dependence of the order parameters in Fig. \ref{fig:gap-temp-px}(a).
At low temperatures, there is weak temperature dependence in $\Delta_x$ and $\Delta_s$,
while they are suppressed as the temperature increases.
We note that the odd-frequency order parameter appears spontaneously for $T<T_{\rm c}$.
In the vicinity of $T_{\rm c}$, $\Delta_x$ show a square root behavior as expected.
On the other hand, $\Delta_s$ show a linear temperature dependence just below $T_{\rm c}$.
To see this point, we also show the temperature dependence of $\Delta_x^2$ in Fig. \ref{fig:gap-temp-px}(a).
We can see that $\Delta_s$ is scaled by $\Delta_x^2$.
From a viewpoint of Ginzburg-Landau theory,
this result implies that there is a third-order term in the free energy for the $s$-wave component as
\begin{align}
F_s = a \Delta_s^2 + b \Delta_s^4 + c \Delta_s \Delta_x^2,
\label{eqn:free-energy}
\end{align}
where $a$, $b$, and $c$ are coefficients.
We assumed real values for the order parameters for simplicity.
In the vicinity of the transition temperature, both order parameters are small
and the fourth order term can be neglected.
Minimizing the free energy with respect to $\Delta_s$, we obtain
\begin{align}
\Delta_s = - \frac{c}{2a} \Delta_x^2.
\label{eqn:Delta-s}
\end{align}
Since $\Delta_x$ shows a square root temperature dependence,
the third-order term in the free energy explains the result
that the $s$-wave component is proportional to $\Delta_x^2$ and shows the linear temperature dependence.

In Fig. \ref{fig:gap-temp-px}(b), we show the coupling constant $(V_s$) dependence of the maximum value of $|\Delta_s(\om,x)|$.
We can see that it shows a linear dependence in the small $V_s$ region.
This is because the $s$-wave order parameter is a product of the induced pair amplitude and the coupling constant $V_s$.
The linear $V_s$ dependence in Fig. \ref{fig:gap-temp-px}(b) implies
that the constant $c$ in Eq. (\ref{eqn:free-energy}) is proportional to $V_s$.

Even though the magnitude of the odd-frequency order parameter is small,
the odd-frequency order parameter is finite as shown in Fig. \ref{fig:gap-temp-px}(b).
The third-order term of the free energy given in Eq. (\ref{eqn:free-energy})
leads to the emergence of the odd-frequency order parameter below $T_{\rm c}$ of the bulk $p$-wave.

\subsection{Repulsive interaction for the odd-frequency channel}
\label{sec:repulsive}

In the previous subsections, we studied an attractive interaction for the odd-frequency order parameter.
In the usual cases, superconducting order parameters are stabilized by attractive interactions.
However, in the present case with the induced odd-frequency pair amplitude, the interaction is not necessary to be attractive.
In this subsection, we consider a case of repulsive interaction ($V_s<0$)
and show that the odd-frequency order parameter exists also in this case.
This can be understood that the sign of the constant $c$ in Eq. (\ref{eqn:Delta-s}) is reversed
in the present repulsive interaction case.
To demonstrate this, we simply use a negative value of the coupling constant in Eq. (\ref{eqn:gap-equation})
for the $s$-wave order parameter.

\begin{figure}[t]
\begin{center}
\includegraphics[width=6.5cm,clip]{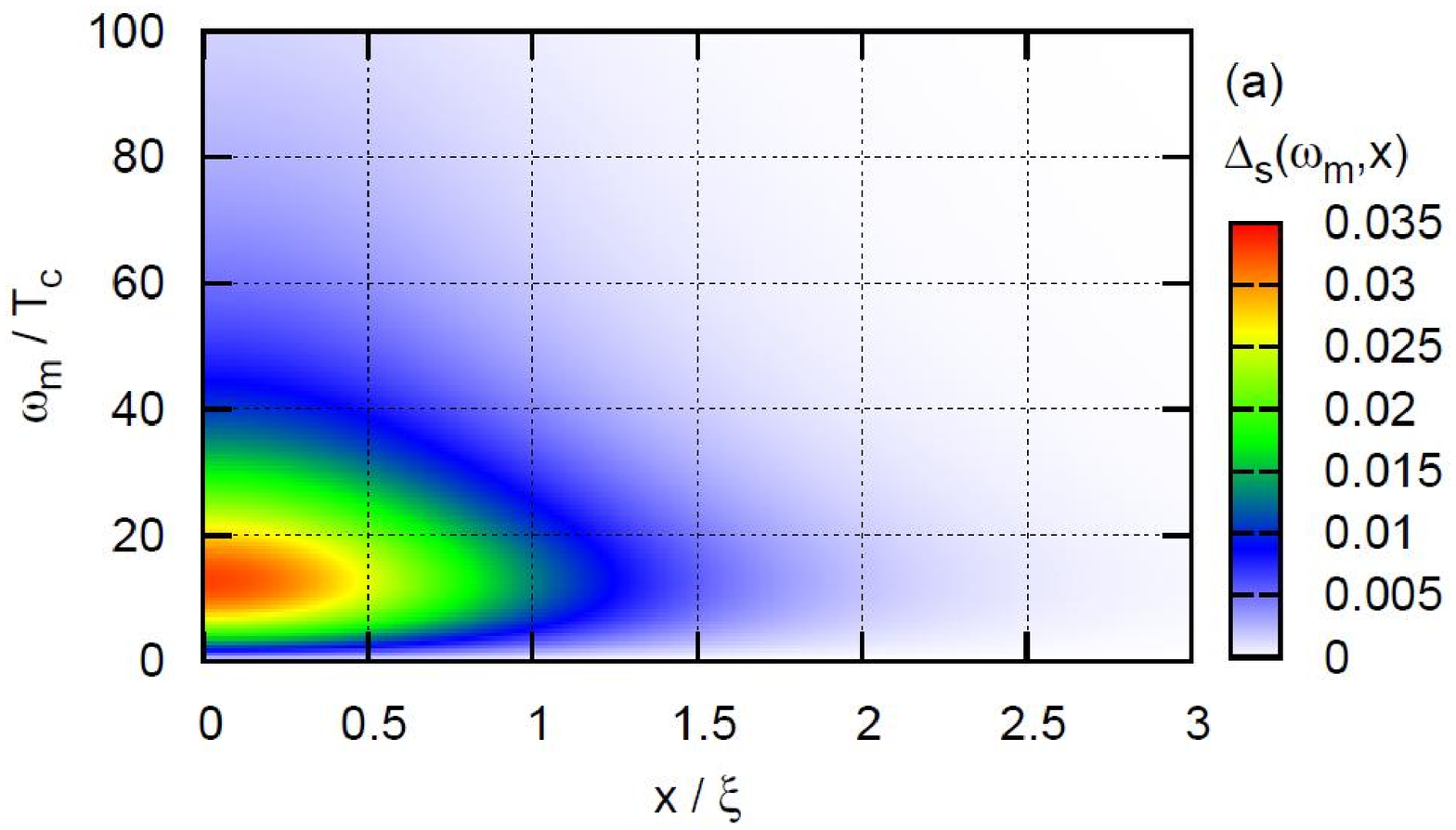}
\includegraphics[width=6.5cm,clip]{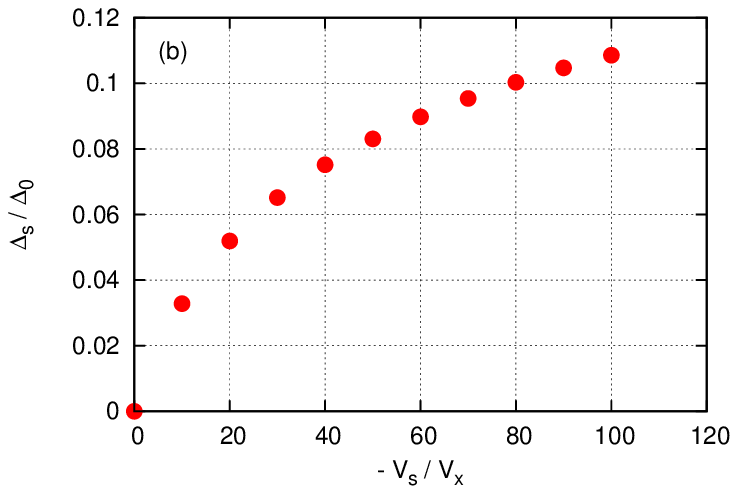}
\end{center}
\caption{
(Color online)
(a) Contour plot of $\Delta_s(\om,x)$.
The coupling constant is chosen as $V_s=-10 V_x$.
(b) Repulsive coupling constant ($V_s<0$) dependence of the maximum value of $|\Delta_s(\om,x)|$ at $x=0$ for $T=0.1 T_{\rm c}$.
}
\label{fig:gap-repulsive}
\end{figure}

In Fig. \ref{fig:gap-repulsive}(a), we show the result for $V_s=-10 V_x$.
The spatial dependence of $\Delta_x(x)$ is basically the same as that for the attractive case shown in Fig. \ref{fig:gap-px}(a).
Concerning the $s$-wave, there is no big difference between the attractive and repulsive cases.
In Fig. \ref{fig:gap-repulsive}(a), we can see that the sign of the order parameter is positive
and is reversed from that of the positive interaction case [see Fig. \ref{fig:gap-px}(d)] reflecting the sign-reversed coupling constant.

Another point is that the magnitude of the order parameter is reduced compared to the attractive case [see Fig. \ref{fig:gap-px}(d)].
In the self-consistent calculation in the repulsive case, the finite pair amplitude gives rise to a positive order parameter.
However, the positive order parameter induces a negative order parameter in the next iteration step
when we solve the gap-equation self-consistently.
Therefore, there is a competition between the pair amplitude and order parameter,
where the former and latter favor the positive and negative sign.
Since the driving force from the pair amplitude is stronger,
the self-consistently determined order parameter becomes positive but its magnitude is strongly reduced from that in the attractive case.
This is also explained by the free energy, since the $a$ term in Eq. (\ref{eqn:Delta-s}) becomes large in the repulsive interaction case.

In Fig. \ref{fig:gap-repulsive}(b), we show the repulsive coupling constant dependence
of the maximum value of $|\Delta_s(\om,x)|$ at $x=0$.
We can see that the slope becomes gradual in the strong repulsive coupling region
which differs from the attractive coupling case shown in Fig. \ref{fig:gap-temp-px}(b).

This type of superconducting order parameter appearing under a repulsive interaction was previously discussed
in a $d_{x^2-y^2}$-wave superconductor near an interface to a normal metal and near a surface.
\cite{Ohashi-1996-2,Matsumoto-1996}
In the latter case, when a surface direction deviates from the symmetric axes, $(1,0,0)$ or $(1,1,0)$,
a finite $s$-wave pair amplitude is induced owing to the symmetry mixing at the surface between the $d_{x^2-y^2}$-wave and $s$-wave.
In this case, the induced pair amplitude has an even-frequency dependence
and it is stabilized as an $s$-wave superconducting order parameter under a finite interaction.
It does not matter if the interaction is attractive or repulsive.
\cite{Ohashi-1996-2,Matsumoto-1996}
We note that the physics of the emergent order parameter under a repulsive interaction is essentially the same
in the $d_{x^2-y^2}$-wave and the present $p$-wave cases.

The repulsive interaction used here is just a theoretical model.
In a realistic case, a short-range Coulomb repulsion can be the candidate for the strong repulsive interaction.

\subsection{$p_x+\ri p_y$-wave case}
\label{subsec:pxy}

\begin{figure}[t]
\begin{center}
\includegraphics[width=6cm,clip]{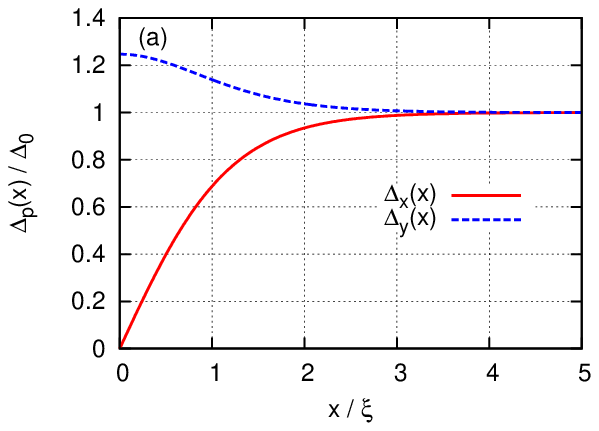}
\includegraphics[width=6cm,clip]{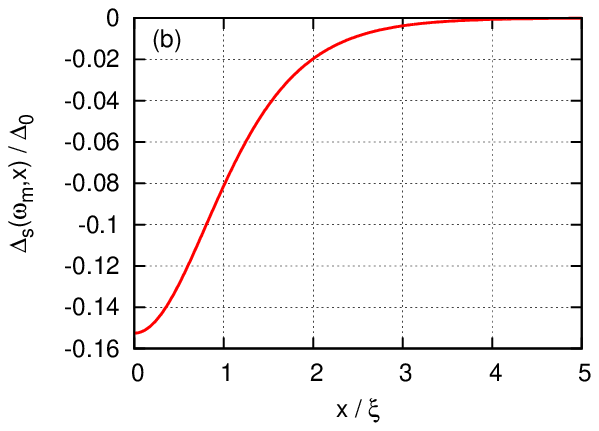}
\includegraphics[width=6cm,clip]{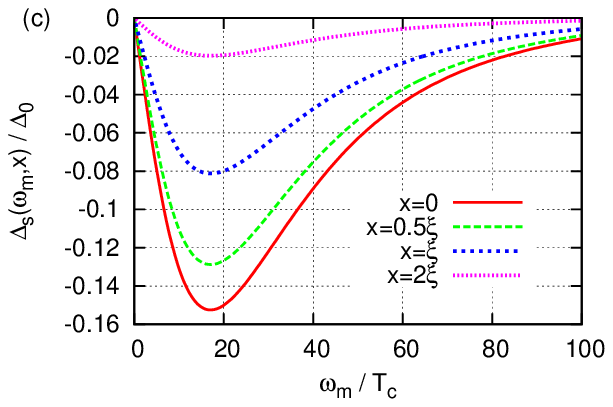}
\includegraphics[width=6.5cm,clip]{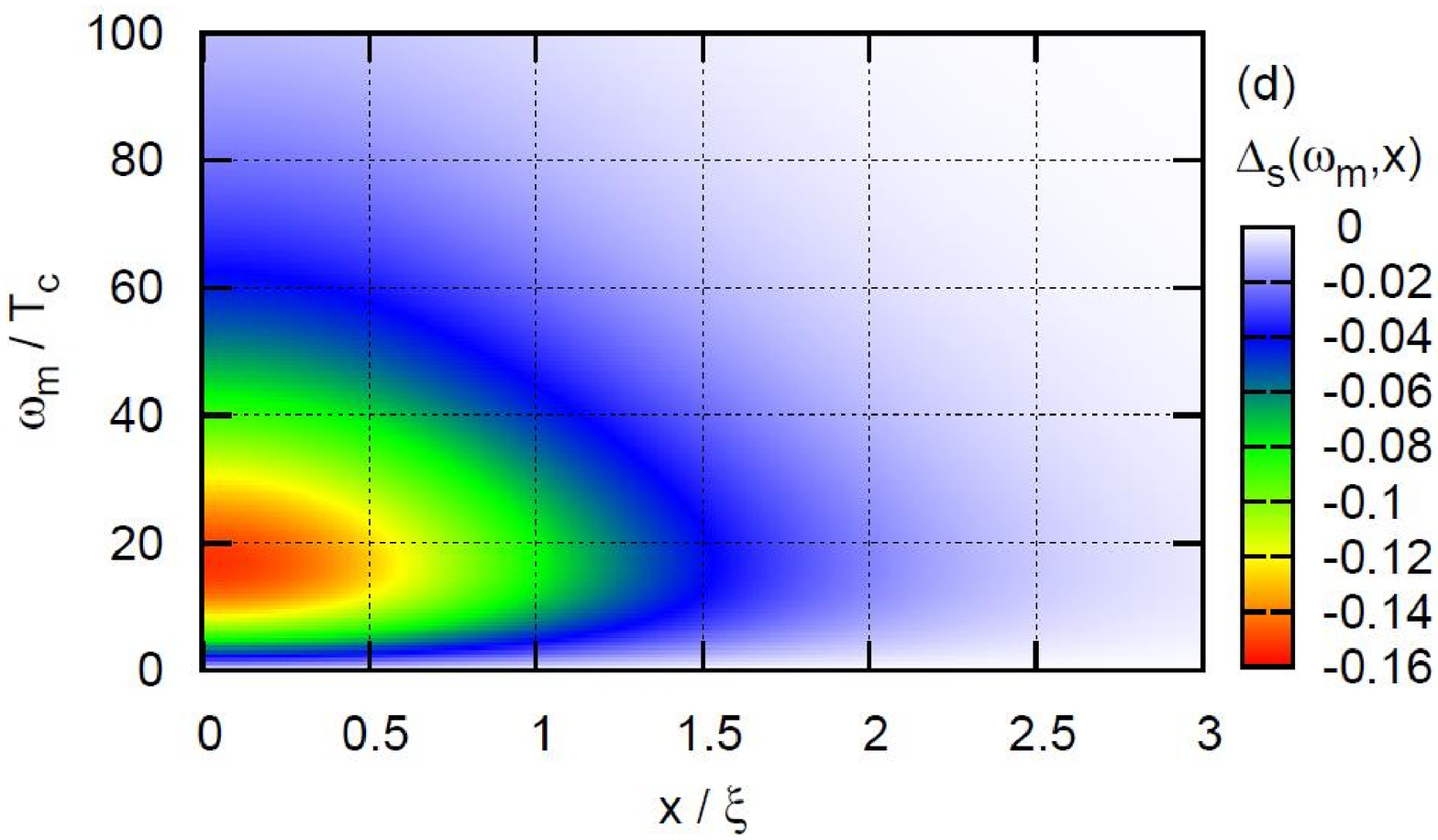}
\end{center}
\caption{
(Color online)
(a) Spatial dependence of the $p$-wave order parameters at $T=0.1 T_{\rm c}$ for $V_x=V_y$.
The parameter is chosen as $\omega_{\rm c}=100 T_{\rm c}$.
(b) Spatial dependence of the $s$-wave order parameter.
The Matsubara frequency is fixed as $\om=17.3 T_{\rm c}$ at which $|\Delta_s(\om,x)|$ has the maximum value.
The set of parameters are chosen as $\omega_{\rm E}=10 T_{\rm c}$ and $V_s=10 V_x$.
(c) Matsubara frequency dependence of the odd-frequency spin triplet $s$-wave order parameter at various positions.
(d) Contour plot of $\Delta_s(\om,x)$.
}
\label{fig:gap-pxy}
\end{figure}

Next, we study the case of the chiral $p_x+\ri p_y$-wave case with $V_x=V_y$.
The transition temperature $T_{\rm c}$ and the solution of the gap equation Eq. (\ref{eqn:gap-equation})
at $T=0$ in the bulk ($x\rightarrow\infty$) are given by
\cite{Matsumoto-1999,Matsumoto-1995-2}
\begin{align}
&T_{\rm c} = \frac{ 2 \re^\gamma \omega_{\rm c} }{\pi} \re^{-\frac{1}{V_x}}, \cr
&\Delta_x(\infty) = \Delta_y(\infty) = 2 \omega_{\rm c} \re^{-\frac{1}{V_x}} \simeq 1.76 T_{\rm c} \equiv \Delta_0.
\end{align}
The coherence length is defined by $\xi = v_{\rm F}/\Delta_0$.
We note that $\Delta_0$ for the $p_x +\ri p_y$-wave is smaller than that for the pure $p_x$-wave case studied in the previous subsection.
Accordingly, $\xi$ for the $p_x +\ri p_y$-wave is larger than that for the $p_x$-wave.

We next show the self-consistently determined order parameters in Fig. \ref{fig:gap-pxy}.
The $p_x$-wave component $\Delta_x(x)$ decreases near the surface, while the $p_y$-wave component $\Delta_y(x)$ is enhanced as expected
[see Fig. \ref{fig:gap-pxy}(a)].
\cite{Matsumoto-1999}
As in the $p_x$-wave case, the odd-frequency order parameter appears near the surface [ see Fig. \ref{fig:gap-pxy}(b)].
It shows the same behavior as that in the $p_x$-wave case as shown in Figs. \ref{fig:gap-pxy}(c) and \ref{fig:gap-pxy}(d).

\subsection{Domain wall ($p_x-\ri p_y|p_x+\ri p_y$) case}

\begin{figure}[t]
\begin{center}
\includegraphics[width=4.5cm,clip]{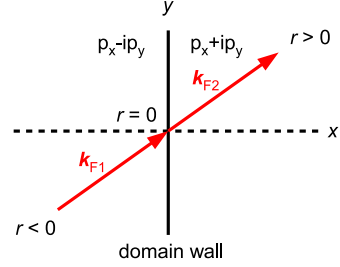}
\end{center}
\caption{
(Color online)
Domain wall between $p_x-\ri p_y$ and $p_x+\ri p_y$ pairing states.
The wavevector is conserved when the quasi-particle passes through the domain wall ($\bkf_1=\bkf_2$).
}
\label{fig:domain}
\end{figure}

In case of the $p_x\pm\ri p_y$-wave superconductors, the $p_x\pm\ri p_y$ states degenerate and break the time-reversal symmetry.
In this case, we can expect a domain wall formed between the two paring states.
There are two types of the domain wall.
One is formed between $p_x-\ri p_y$ and $p_x+\ri p_y$,
while the other is formed between $-p_x+\ri p_y$ and $-p_x+\ri p_y$.
We note that the former domain wall is energetically favorable and is stabilized against the latter case.
\cite{Volovik,Sigrist-1989,Sigrist-1991}
Near such domain wall, bound states exist owing to the sign change of the order parameter of the $p_y$-wave
after passing through the domain wall.
In this case, we can also expect that the odd-frequency order parameter appears.

We consider a domain wall shown in Fig. \ref{fig:domain}.
We assume that there is no scattering at the domain wall and that the quasi-particle travels along a straight line.
Since the odd-frequency amplitude is induced by the sign-changing component ($p_y$-wave),
the odd-frequency order parameter is pure imaginary.
In case of the domain wall $p_x-\ri p_y|p_x+\ri p_y$, the induced odd-frequency amplitude has a spin-triplet $d_{xy}$-wave symmetry.
We will discuss this point later in \S \ref{subsec:symmetry}.
Then, the order parameters are expressed as
\begin{align}
&\Delta_1(\bkf,\om,x) \cr
&= \Delta_x(x) \cos\tk + \ri \Delta_y(x) \sin\tk + \ri \Delta_d(\om,x) \sin 2\tk, \cr
&\Delta_2(\bkf,\om,x) \cr
&= \Delta_x(x) \cos\tk - \ri \Delta_y(x) \sin\tk + \ri \Delta_d(\om,x) \sin 2\tk. \cr
\end{align}
Here, $\ri\Delta_d(\om,x) \sin 2\tk$ represents the $d_{xy}$-wave order parameter.
To represent the $d_{xy}$-wave symmetry, we use $\sin2\tk$ function for simplicity.
We can solve the quasi-classical Green's function along the quasi-classical trajectory as in the scattering at the surface case.
In case of the domain wall, the gap-equation is altered as
\begin{align}
&\left(
  \begin{matrix}
    \Delta_x(x) \cr
    \Delta_y(x) \cr
    \Delta_d(\ol,x) \cr
  \end{matrix}
\right)
= T \sum_{0 < \om < \omega_{\rm c}} \int_0^{\frac{\pi}{2}} \rd \tk \cr
&~~~\times
\left(
  \begin{matrix}
    2 V_x \cos \tk f_x(\tk,\om,x) \cr
    -\ri 2 V_y \sin \tk f_y(\tk,\om,x) \cr
    -\ri 2 V_d V_-(\ol,\om) \sin 2\tk f_d(\tk,\om,x)
  \end{matrix}
\right),
\end{align}
where
\begin{align}
&f_x(\tk,\om,x) = f(\tk,\om,x) + \fbar^*(\tk,\om,x) \cr
&~~~~~~~~~~~~~~~~~
                + f(-\tk,\om,x) + \fbar^*(-\tk,\om,x), \cr
&f_y(\tk,\om,x) = f(\tk,\om,x) + \fbar^*(\tk,\om,x) \cr
&~~~~~~~~~~~~~~~~~
                - f(-\tk,\om,x) - \fbar^*(-\tk,\om,x), \cr
&f_d(\tk,\om,x) = f(\tk,\om,x) - \fbar^*(\tk,\om,x) \cr
&~~~~~~~~~~~~~~~~~
                - f(-\tk,\om,x) + \fbar^*(-\tk,\om,x).
\end{align}
The following relations were used to derive the gap equation in the present case:
\begin{align}
&f(\tk+\pi,-\om,x) = - f(\tk,\om,x), \cr
&f(\tk,-\om,x) = \fbar^*(\tk,\om,x).
\end{align}
To demonstrate the emergent odd-frequency $d_{xy}$-wave order parameter, we simply use $V_-(\ol,\om)$ defined by Eq. (\ref{eqn:V-odd})
as the frequency dependence of the effective interaction for the $d_{xy}$-wave.

\begin{figure}[t]
\begin{center}
\includegraphics[width=6cm,clip]{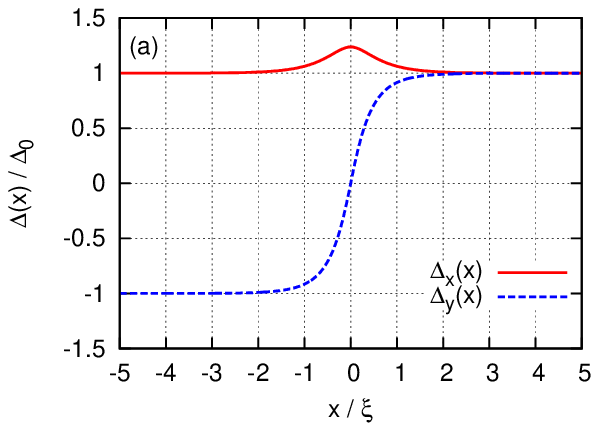}
\includegraphics[width=6cm,clip]{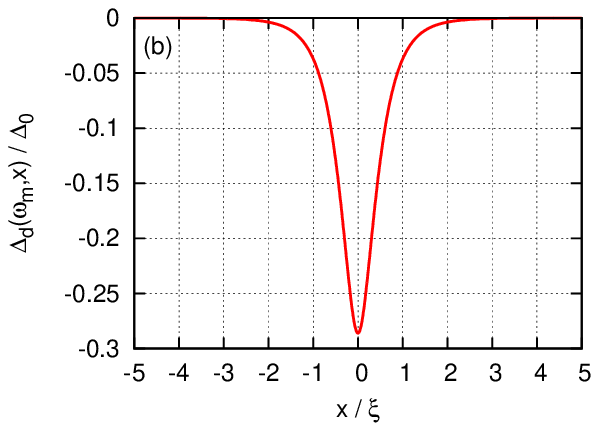}
\includegraphics[width=6cm,clip]{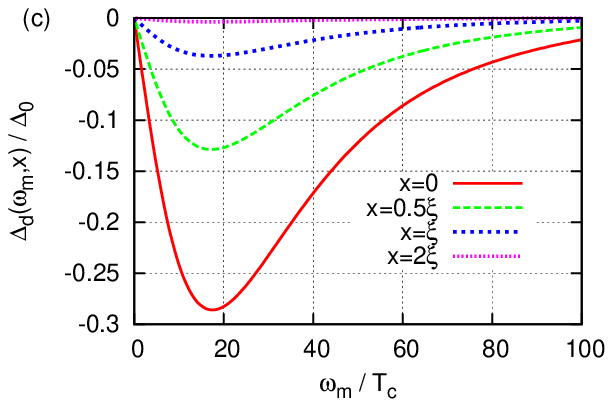}
\includegraphics[width=6.5cm,clip]{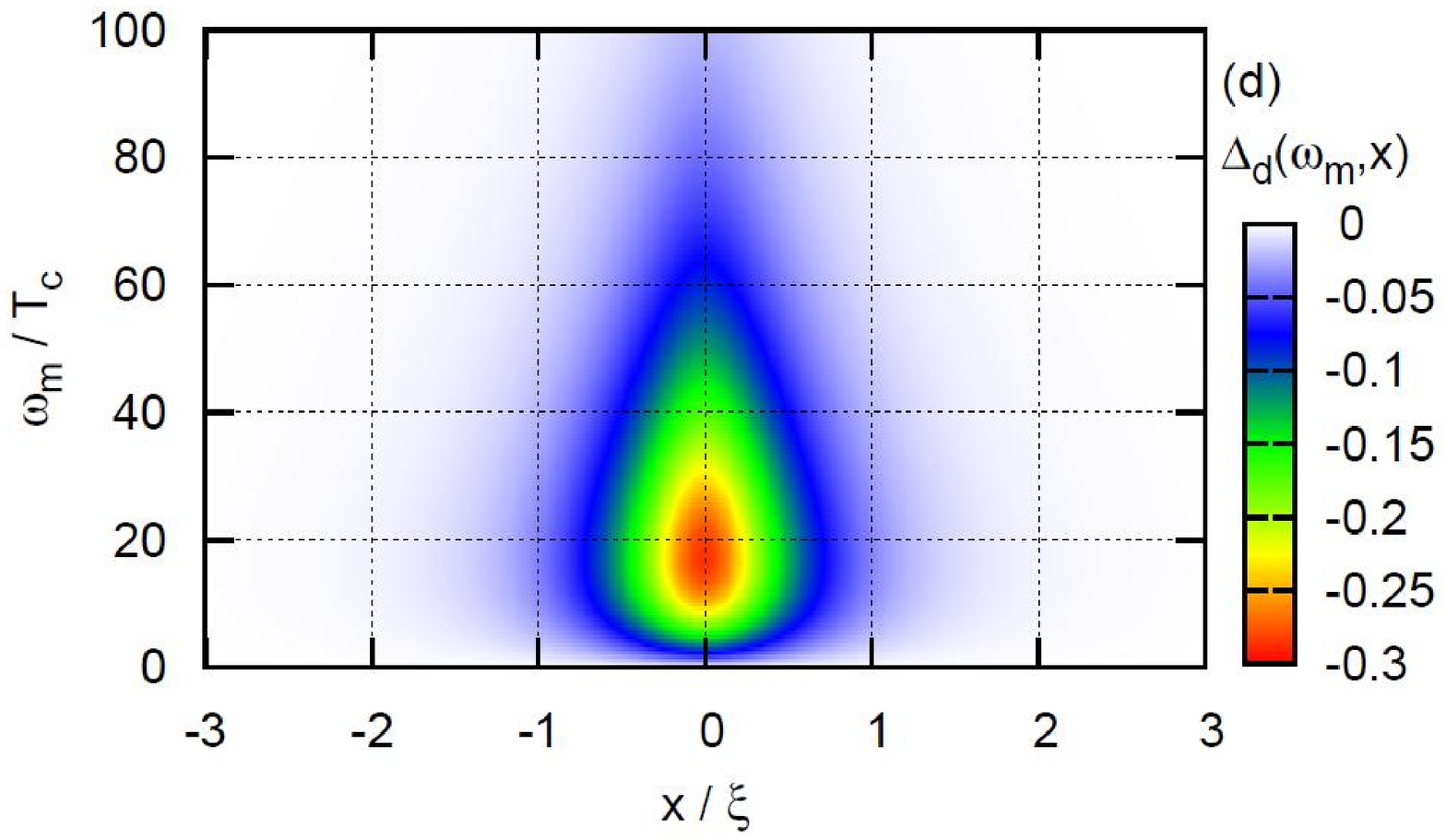}
\end{center}
\caption{
(Color online)
(a) Spatial dependence of the $p$-wave order parameters at $T=0.1 T_{\rm c}$ for $V_x=V_y$.
The parameter is chosen as $\omega_{\rm E}=10 T_{\rm c}$.
(b) Spatial dependence of the $d_{xy}$-wave order parameter.
The Matsubara frequency is fixed as $\om=17.3 T_{\rm c}$ at which $|\Delta_d(\om,x)|$ has the maximum value.
The set of parameters are chosen as $\omega_{\rm E}=10 T_{\rm c}$ and $V_d=10 V_x$.
(c) Matsubara frequency dependence of the odd-frequency spin-triplet $d_{xy}$-wave order parameter at various positions.
We note that $\Delta_d(\om,x)$ is an even function with respect to $x$.
(d) Contour plot of $\Delta_d(\om,x)$.
}
\label{fig:gap-pxy-domain}
\end{figure}

In Fig. \ref{fig:gap-pxy-domain}, we show the result.
In the domain wall case, $\Delta_y(x)$ is suppressed near the boundary,
while $\Delta_x(x)$ is enhanced as shown in Fig. \ref{fig:gap-pxy-domain}(a).
As in the surface case, a finite $d_{xy}$-wave order parameter appears with an odd-frequency dependence
as shown in Figs. \ref{fig:gap-pxy-domain}(b) and \ref{fig:gap-pxy-domain}(c).
In Fig. \ref{fig:gap-pxy-domain}(d), we show the contour plot of $\Delta_d(\om,x)$.

\section{Emergent Odd-Frequency Order Parameter and Mid-Gap Bound State}
\label{sec:pair}

It is well known that surface breaks the unconventional superconducting order parameter and induces surface bound state.
For instance, in Copper oxide high temperature superconductors,
(1,1,0) surface breaks the $d_{x^2-y^2}$-wave order parameter and induces the mid-gap surface bound state.
\cite{Hu,Kashiwaya,Tanaka-1995,Matsumoto-1995-1,Nagato-1995,Ohashi-1996-1}
When the time-reversal symmetry is broken, the energy of the mid-gap (zero-energy) bound state is shifted to a finite energy
and it carries a surface current to induce a spontaneous magnetic field localized in the vicinity of the surface.
\cite{Matsumoto-1995-2,Matsumoto-1995-3,Matsumoto-1996,Covington,Fogelstrom,Honerkamp,Matsumoto-1999,Furusaki}

It was previously reported that the induced odd-frequency pair amplitude is related closely to the existence of the bound state.
\cite{Tanaka-2007-2,Tanaka-2007-3,Tanaka-2012}
In this section, we examine this point in terms of the quasi-particle wavefunction by solving the Andreev equation.

\subsection{Quasi-classical wavefunction}

\subsubsection{Andreev equation}

Let us begin with the following Andreev equation for unconventional superconductors:
\cite{Andreev,Bruder}
\begin{align}
\left(
  \begin{matrix}
    - \ri\bvf\cdot\nabla & \Delta(\bkf,\br) \cr
    \Delta(\bkf,\br) & \ri\bvf\cdot\nabla
  \end{matrix}
\right)
\left(
  \begin{matrix}
    u_E(\br) \cr
    v_E(\br)
  \end{matrix}
\right)
= E
\left(
  \begin{matrix}
    u_E(\br) \cr
    v_E(\br)
  \end{matrix}
\right).
\label{eqn:Andreev-main}
\end{align}
Here, we focus on the $p_x$-wave state with a real value of the order parameter.
The Andreev equation enables us to investigate the quasi-particle states in terms of the slowly-varying function
of the order of the superconducting coherence length, where the rapid oscillation of the order of the Fermi wave length is averaged out.
The derivation of the Andreev equation is given in Appendix \ref{appendix:Andreev}.
There are positive ($E>0$) and negative ($E<0$) solutions in Eq. (\ref{eqn:Andreev-main}).
They correspond to the particle and hole solutions, respectively, and are not independent each other.
There is the following relation between these solutions:
\begin{align}
\left(
  \begin{matrix}
    u_{-E}(\br) \cr
    v_{-E}(\br)
  \end{matrix}
\right)
\leftrightarrow
\left(
  \begin{matrix}
    -v_E(\br) \cr
    u_E(\br)
  \end{matrix}
\right).
\label{eqn:E-pm}
\end{align}
Summing all energy eigenvalues, the field operators for a fixed $\bkf$ are written as
\begin{align}
&\left(
  \begin{matrix}
    \psi_{\bkf\uparrow}(\br) \cr
    \psi_{-\bkf\downarrow}^\dagger(\br)
  \end{matrix}
\right)
=
\left(
  \begin{matrix}
    u_0 \cr
    v_0
  \end{matrix}
\right) \gamma_0 \cr
&~~~~~~~~~~~~
+ \sum_{E>0}
\left[
\left(
  \begin{matrix}
    u_E \cr
    v_E
  \end{matrix}
\right) \gamma_{E\uparrow}
+
\left(
  \begin{matrix}
    -v_E \cr
     u_E
  \end{matrix}
\right) \gamma_{E\downarrow}^\dagger
\right],
\label{eqn:psi-kf}
\end{align}
where the definition of the field operators are given by Eq. (\ref{eqn:psi-uv}).
In Eq. (\ref{eqn:psi-kf}), the first term is for the mid-gap (zero-energy) state.
The second and third terms are for the particle and hole solutions, respectively.
We introduced two kinds of fermions, $\gamma_{E\uparrow}$ and $\gamma_{E\downarrow}$.
In case of the mid-gap state, however, we note that there is only one energy eigenstate for each $\bkf$.
For a fixed $\bkf$, the Hamiltonian is expressed in the following diagonal form
with the $\gamma$ fermions for superconducting quasi-particles:
\begin{align}
\mathcal{H}_\bkf =
  0 \times \gamma_0^\dagger \gamma_0
+ \sum_{E > 0} E \left( \gamma_{E \uparrow}^\dagger \gamma_{E \uparrow}
                      + \gamma_{E \downarrow}^\dagger \gamma_{E \downarrow} \right).
\end{align}

\subsubsection{Solution of Andreev equation}

In this subsection, we focus on the $p_x$-wave superconductor with a specular surface at $x=0$.
Since the spatial dependence of the order parameter is irrelevant to understand the mid-gap state,
we assume a constant order parameter to solve the Andreev equation.
As discussed in Appendix \ref{appendix:boundary-condition}, we can solve the Andreev equation along the quasi-classical trajectory
as shown in Fig. \ref{fig:boundary}.
It is convenient to introduce the following functions:
\begin{align}
f_\pm(r) = u(r) \pm i v(r).
\end{align}
Here, $r$ is the coordinate along the trajectory.
We dropped the subscript `$E$' in $u$, $v$, and $f_\pm$ for convenience.
The Andreev equation is then expressed as
\cite{Takayama}
\begin{align}
E f_\pm(r) = -\ri \vf \frac{\partial}{\partial r} f_\mp(r) \pm \ri \Delta(\bkf,r) f_\mp(r).
\label{eqn:f-pm}
\end{align}
We differentiate Eq. (\ref{eqn:f-pm}) with respect to $r$ and obtain
\cite{Takayama}
\begin{align}
\left[
  \vf^2 \frac{\partial^2}{\partial r^2} + E^2 - \Delta^2(\bkf,r) \pm \vf \frac{\partial}{\partial r} \Delta(\bkf,r)
\right] f_\pm(r) = 0.
\label{eqn:f-pm-1}
\end{align}
Along the quasi-classical trajectory, the order parameter is uniform,
however, its sign suddenly changes from negative to positive at $r=0$ [see Fig. \ref{fig:boundary}(b)].
Therefore, the derivative of the order parameter is replaced as
$\partial\Delta(\bkf,r)/\partial r = 2 |\Delta_\bkf| \delta(r)$.
Here, $\Delta_\bkf$ is the uniform order parameter for $\bkf$.
$\delta(r)$ is the Dirac's delta-function.
Introducing the coherence length $\xi=\vf/|\Delta_\bkf|$, we can rewrite Eq. (\ref{eqn:f-pm-1}) as
\begin{align}
\left[
  - \xi^2 \frac{\partial^2}{\partial r^2} \mp 2 \xi \delta(r)
\right] f_\pm(r) = \left( \varepsilon^2 - 1 \right) f_\pm(r).
\label{eqn:sch}
\end{align}
Here, $\varepsilon$ is a dimensionless energy defined by $\varepsilon=E/|\Delta_\bkf|$.
We note that Eq. (\ref{eqn:sch}) has the same form of a Schr\"{o}dinger equation with a delta-function potential.
Since the potential is attractive, there is one bound state at $\varepsilon=0$ for $f_+(r)$, while there is no bound state for $f_-(r)$.
The bound state solution is given by
\begin{align}
f_+^{\rm B}(r) = \sqrt{ \frac{2}{\xi}} \re^{-\frac{|r|}{\xi}},~~~~~~f_-^{\rm B}(r) = 0.
\end{align}
In case of the continuum state ($|\varepsilon|>1$), we classify the solution into two types following Takayama {\it et al}.
\cite{Takayama}
\begin{align}
&f_+^{(1)}(-r) = f_+^{(1)}(r),~~~~~~f_-^{(1)}(-r) = - f_-^{(1)}(r), \cr
&f_+^{(2)}(-r) = - f_+^{(2)}(r),~~~~~~f_-^{(2)}(-r) = f_-^{(2)}(r).
\end{align}
The explicit form of the solution is given by
\begin{align}
&f_+^{(1)}(r) = \sqrt{ \frac{2}{L} } \frac{1}{1+\ri \xi k} \left( \xi k\cos{kr} - \sin{k|r|} \right), \cr
&f_-^{(1)}(r) = \ri \sqrt{ \frac{2}{L} } \sqrt{ \frac{1-\ri \xi k}{1+\ri \xi k} } \sin{kr}, \cr
&f_+^{(2)}(r) = \ri \sqrt{ \frac{2}{L} } \sqrt{ \frac{1-\ri \xi k}{1+\ri \xi k} } \sin{kr}, \cr
&f_-^{(2)}(r) = \sqrt{ \frac{2}{L} } \frac{1}{1+\ri \xi k} \left( \xi k\cos{kr} + \sin{k|r|} \right),
\end{align}
where the energy eigenvalue is given by $\varepsilon=\sqrt{1+(\xi k)^2}$.
The positive wave numbers $k>0$ are sufficient for the solution,
since the wavefunctions are identical under $k\rightarrow -k$ except for the phase factor.
The normalization condition for $f_\pm$ is chosen as
\cite{Takayama,note-L}
\begin{align}
\int_{-\frac{L}{2}}^{\frac{L}{2}} \rd r \left[ |f_+(r)|^2 + |f_-(r)|^2 \right] = 2.
\end{align}
The corresponding $u$ and $v$ are given by
\begin{align}
&u_0(r) = \frac{1}{\sqrt{2\xi}} \re^{-\frac{|r|}{\xi}},
~~~~~~
v_0(r) = -\ri \frac{1}{\sqrt{2\xi}} \re^{-\frac{|r|}{\xi}}, \cr
&u^{(1)}(r) = \frac{1}{2} \sqrt{ \frac{2}{L} } \frac{1}{1+\ri \xi k} \cr
&~~~~~~\times
              \left[ \xi k\cos{kr'} - \sin{k|r'|} + \ri \sqrt{1+(\xi k)^2} \sin{kr'} \right], \cr
&v^{(1)}(r) = -\ri \frac{1}{2} \sqrt{ \frac{2}{L} } \frac{1}{1+\ri \xi k} \cr
&~~~~~~\times
                   \left[ \xi k\cos{kr'} - \sin{k|r'|} - \ri \sqrt{1+(\xi k)^2} \sin{kr'} \right], \cr
&u^{(2)}(r) = \frac{1}{2} \sqrt{ \frac{2}{L} } \frac{1}{1+\ri \xi k} \cr
&~~~~~~\times
              \left[ \xi k\cos{kr'} + \sin{k|r'|} + \ri \sqrt{1+(\xi k)^2} \sin{kr'} \right], \cr
&v^{(2)}(r) = \ri \frac{1}{2} \sqrt{ \frac{2}{L} } \frac{1}{1+\ri \xi k} \cr
&~~~~~~\times
                  \left[ \xi k\cos{kr'} + \sin{k|r'|} - \ri \sqrt{1+(\xi k)^2} \sin{kr'} \right]. \cr
\label{eqn:solution}
\end{align}
Here, $u_0$ and $v_0$ are wavefunctions for the mid-gap bound state,
while $u^{(i)}$ and $v^{(i)}$ ($i=1,2$) are for the continuum states.
We note that $u$ and $v$ satisfy the following normalization condition:
\begin{align}
\int_{-\frac{L}{2}}^{\frac{L}{2}} \rd r \left[ |u(r)|^2 + |v(r)|^2 \right] = 1.
\end{align}

\subsection{Superconducting pair amplitude}

\subsubsection{General description}

The superconducting pair amplitude is defined by the following form:
\begin{align}
&F_{\uparrow\downarrow}(\br,\br',\tau) = - \langle T_\tau \psi_\uparrow(\br,\tau) \psi_\downarrow(\br') \rangle \cr
&~~~
= - \langle \psi_\uparrow(\br,\tau) \psi_\downarrow(\br') \rangle \theta(\tau)
   + \langle \psi_\downarrow(\br') \psi_\uparrow(\br,\tau) \rangle \theta(-\tau). \cr
\label{eqn:F}
\end{align}
Here, $T_\tau$ represents the time-order operator with respect to the imaginary time $\tau$ and $\theta(\tau)$ is the step function.
Substituting Eq. (\ref{eqn:psi-k}) into Eq. (\ref{eqn:F}), we obtain
\begin{align}
&F_{\uparrow\downarrow}(\br,\br',\tau) \cr
&= - \sum_{\bkf\bkf'} \re^{\ri\bkf\cdot\br} \re^{-\ri\bkf'\cdot\br'}
                       \langle T_\tau \psi_{\bkf\uparrow}(\br,\tau) \psi_{-\bkf'\downarrow}(\br') \rangle \cr
&\simeq - \sum_\bkf \re^{\ri\bkf\cdot(\br-\br')}
             \left\langle T_\tau
                 \left( 1 + \frac{\bx}{2}\cdot\nabla_R \right) \psi_{\bkf\uparrow}(\bR,\tau) \right. \cr
&~~~~~~~~~~~~~~~~~~~~~~~~~~\times                 
             \left.                 
                 \left( 1 - \frac{\bx}{2}\cdot\nabla_R \right) \psi_{-\bkf\downarrow}(\bR)
             \right\rangle \cr
&\simeq \sum_\bkf \re^{\ri\bkf\cdot(\br-\br')} F_{\bkf\uparrow\downarrow}(\bR,\tau),
\end{align}
where $\bR$ and $\bx$ represent the center of mass and relative coordinate of the Cooper pair, respectively.
$F_{\bkf\uparrow\downarrow}(\br,\tau)$ is defined by
\begin{align}
F_{\bkf\uparrow\downarrow}(\br,\tau) = - \left\langle T_\tau \psi_{\bkf\uparrow}(\br,\tau) \psi_{-\bkf\downarrow}(\br) \right\rangle
\end{align}
as the Fourier transformed pair amplitude.
We note that $\br$ in $F_{\bkf\uparrow\downarrow}(\br,\tau)$ corresponds to the center of mass coordinate of the Cooper pair.
The Fourier transformed pair amplitude with respect to $\tau$ is defined by
\begin{align}
F_{\bkf\uparrow\downarrow}(\br,\ri\om)
= \int_0^\beta \rd \tau \re^{\ri\omega_m\tau} F_{\bkf\uparrow\downarrow}(\br,\tau).
\label{eqn:F-kf}
\end{align}
Here, $\om$ is the fermionic Matsubara frequency and $\beta=1/T$.

In the following discussion, we express $F_{\bkf\uparrow\downarrow}(\br,\ri\om)$
with the energy eigenstates of the Andreev equation and examine what state forms the odd-frequency pair amplitude.
For this purpose, we express the pair amplitude by using the wavefunctions $u$ and $v$.
Substituting Eq. (\ref{eqn:psi-kf}) into Eq. (\ref{eqn:F-kf}), we obtain
\begin{align}
&F_{\bkf\uparrow\downarrow}(\br,\ri\om)
= - \int_0^\beta \rd \tau \re^{\ri\om\tau} \left\langle \psi_{\bkf\uparrow}(\br,\tau) \psi_{-\bkf\downarrow}(\br) \right\rangle \cr
&= - \int_0^\beta \rd \tau \re^{\ri\om\tau}
\left\{
  u_0(\br) v_0^*(\br) \langle \gamma_0 \gamma_0^\dagger \rangle
\right. \cr
&~~~~~~~~~~~~~~~
\left.  
  + \sum_{E>0}
    \left[ u_E(\br) v_E^*(\br) \re^{-E\tau} \langle \gamma_{E\uparrow} \gamma_{E\uparrow}^\dagger \rangle \right. \right. \cr
&~~~~~~~~~~~~~~~~~~~~~~
\left. \left. 
         - u_E^*(\br) v_E(\br) \re^{E\tau} \langle \gamma_{E\downarrow}^\dagger \gamma_{E\downarrow} \rangle
    \right]
\right\} \cr
&= \frac{1}{\ri\om} u_0(\br) v_0^*(\br)
 + \sum_{E>0} 2 {\rm Re} \left[ \frac{1}{\ri\om - E} u_E(\br) v_E^*(\br) \right].
\label{eqn:pair}
\end{align}
Here we assumed a low temperature limit.
The first term is from the mid-gap bound state, while the second term is from the continuum states.
It is clear that the first term has an odd-frequency dependence.
Equation (\ref{eqn:pair}) clearly indicates that the mid-gap bound state generates the odd-frequency pair amplitude.
It is the key to understand the emergent odd-frequency order parameter.

\subsubsection{Pair amplitude near the boundary}

We discuss next the pair amplitude in terms of the solution of the Andreev equation.
Substituting Eq. (\ref{eqn:solution}) into Eq. (\ref{eqn:pair}) and summing up the two types of solutions,
we can divide the pair amplitude into the even- and odd-frequency components as
\begin{align}
F_{\bkf\uparrow\downarrow}(r,\ri\om)
= F_{\bkf\uparrow\downarrow}^{\rm even}(r,\ri\om) + F_{\bkf\uparrow\downarrow}^{\rm odd}(r,\ri\om).
\end{align}
Here, $F_{\bkf\uparrow\downarrow}^{\rm even}(r,\ri\om)$ and $F_{\bkf\uparrow\downarrow}^{\rm odd}(r,\ri\om)$
are the even- and odd-frequency components defined by
\begin{align}
&F_{\bkf\uparrow\downarrow}^{\rm even}(r,\ri\om)
= - \frac{4}{L} \sum_{k>0} \frac{|\Delta_\bkf|{\rm sgn}(r)}{\om^2 + \Delta_\bkf^2 + (\vf k)^2} \sin^2{kr}, \cr
&F_{\bkf\uparrow\downarrow}^{\rm odd}(r,\ri\om)
= \frac{1}{2\om\xi} \re^{-\frac{2|r|}{\xi}} \cr
&- \om \frac{2}{L} \sum_{k>0} \frac{|\Delta_\bkf|}{\om^2 + \Delta_\bkf^2 + (\vf k)^2}
                              \frac{\vf k}{\Delta_\bkf^2+(\vf k)^2} \sin{2k|r|}.
\end{align}
The even-frequency component is an odd function with respect to $r$ reflecting the symmetry of the $p_x$-wave.
We note that only the continuum states contribute to the even-frequency component.
On the other hand, the odd-frequency component is an even function with respect to $r$.
This indicates that it has an $s$-wave symmetry.
In the odd-frequency component, both of the bound and the continuum states contribute to the pair amplitude.
In the following subsections, we study the two frequency components separately.

\subsubsection{Odd-frequency pair amplitude}

We first replace the summation over $k$ with its integral.
The pair amplitude is then expressed as
\begin{align}
&F_{\bkf\uparrow\downarrow}^{\rm odd}(r,\ri\om) = \frac{1}{2\om\xi} \re^{-\frac{2|r|}{\xi}} \cr
&
-\om \frac{1}{\pi} \int_0^\infty \rd k
                       \frac{|\Delta_\bkf|}{\om^2 + \Delta_\bkf^2 + (\vf k)^2} \frac{\vf k}{\Delta_\bkf^2+(\vf k)^2} \sin{2k|r|} \cr
&= \ri \frac{1}{2\vf} \int_{-\infty}^\infty \rd z \frac{1}{\ri\om-z} \cr
&~~~~~~~~~~~~~~~~~~\times \frac{1}{-\pi}
  {\rm Im}
  \left[
    \frac{|\Delta_\bkf|}{z+\ri\delta} \re^{-2\sqrt{(-\ri z + \delta)^2+\Delta_\bkf^2}|r|/\vf}
  \right] \cr
&= \ri \frac{1}{2\vf}
    \frac{|\Delta_\bkf|}{\ri\om} \re^{-2\sqrt{\om^2+\Delta_\bkf^2}|r|/\vf},
\end{align}
where we introduced $z=\sqrt{\Delta_\bkf^2+(\vf k)^2}$
and used the Lehmann representation with an infinitesimal small positive number $\delta$.
In terms of the $x(\ge 0)$ coordinate, where $x=|\vfx||r|/\vf$, we obtain
\begin{align}
F_{\bkf\uparrow\downarrow}^{\rm odd}(x,\ri\om)
= \frac{1}{2\vf} \frac{|\Delta_\bkf|}{\om} \re^{-2\sqrt{\om^2+\Delta_\bkf^2}x/|\vfx|}.
\label{eqn:F-odd}
\end{align}
We can see that the odd-frequency pair amplitude has the $s$-wave symmetry, i.e.,
$F_{\bkf\uparrow\downarrow}^{\rm odd}(x,\ri\om)$ is symmetric with respect to
${\kf}_x\leftrightarrow -{\kf}_x$ and ${\kf}_y\leftrightarrow -{\kf}_y$.
The pair amplitude decreases exponentially as $x$ increases,
reflecting the fact that it is formed mainly by the mid-gap bound state.
We note that Eq. (\ref{eqn:F-odd}) is consistent with the result of the odd-frequency pair amplitude
obtained by the quasi-classical Green's function ($g_1$ component) given in Eq. (\ref{eqn:g-constant}).

\subsubsection{Even-frequency pair amplitude}

The even-frequency pair amplitude is rewritten as
\begin{align}
&F_{\bkf\uparrow\downarrow}^{\rm even}(r,\ri\om)
= - \frac{1}{\pi} \int_0^\infty \rd k
                  \frac{|\Delta_\bkf|{\rm sgn}(r)}{\om^2 + \Delta_\bkf^2 + (\vf k)^2} 2\sin^2{kr} \cr
&= \frac{1}{2\vf} \int_{-\infty}^\infty \rd z \frac{1}{\ri\om-z} \cr
&\times \frac{1}{-\pi}
   {\rm Im} \left[ \frac{-|\Delta_\bkf|{\rm sgn}(r)}{\sqrt{(-\ri z + \delta)^2+\Delta_\bkf^2}} \right. \cr
&~~~~~~~~~~~~~~~~~~~\times \left.
              \left( 1 - \re^{-2{\sqrt{(-\ri z + \delta)^2+\Delta_\bkf^2}|r|}/{\vf}} \right)
            \right] \cr
&= \frac{1}{2\vf} \frac{-|\Delta_\bkf|{\rm sgn}(r)}{\sqrt{\om^2+\Delta_\bkf^2}}
                  \left( 1 - \re^{-2{\sqrt{\om^2+\Delta_\bkf^2}|r|}/{\vf}} \right).
\end{align}
In terms of the $x$ coordinate, where $|\Delta_\bkf|{\rm sgn}(r)\rightarrow \Delta_\bkf$, we obtain
\begin{align}
&F_{\bkf\uparrow\downarrow}^{\rm even}(x,\ri\om) \cr
&= \frac{1}{2\vf} \frac{-\Delta_\bkf}{\sqrt{\om^2+\Delta_\bkf^2}}
                 \left( 1 - \re^{-2{\sqrt{\om^2+\Delta_\bkf^2}x}/{|\vfx|}} \right). \cr
\label{eqn:F-even}
\end{align}
We can see that the even-frequency pair amplitude has the $p_x$-wave symmetry.
We note that Eq. (\ref{eqn:F-even}) is consistent with the result of the even-frequency pair amplitude
obtained by the quasi-classical Green's function ($g_2$ component) given in Eq. (\ref{eqn:g-constant}).

\subsubsection{$p_x+\ri p_y$-wave case}

In the chiral $p_x+\ri p_y$-wave case, the bound state energy is located at $E=\Delta_y(\bkf)$ for each ${\kf}_y$
\cite{Honerkamp,Matsumoto-1999}
and there is no mid-gap bound state except for $\Delta_y(\bkf)=0$.
For ${\kf}_y>0$, the bound state energy is positive, while it is negative for ${\kf}_y<0$.
Thus, the bound states are not symmetric with respect to ${\kf}_y$ and they carry a finite surface current
reflecting the broken time-reversal symmetry.
\cite{Matsumoto-1999,Matsumoto-1995-3}

When the time-reversal symmetry is broken, the relation given in Eq. (\ref{eqn:E-pm}) does not hold
and we have to solve both the particle and hole solutions.
Let us consider the bound state contribution.
In the present $p_x+\ri p_y$-wave case, the bound state contribution in Eq. (\ref{eqn:pair}) is replaced as
\begin{align}
\frac{1}{\ri\om} u_0(r) v_0^*(r)
\rightarrow
\frac{1}{\ri\om-\Delta_y(\bkf)} u^{\rm B}(r) [v^{\rm B}(r)]^*,
\label{eqn:odd-f-pxy}
\end{align}
where $u^{\rm B}(r)$ and $v^{\rm B}(r)$ are wavefunctions for the bound state at $E=\Delta_y(\bkf)$.
For constant $\Delta_x(\bkf)$ and $\Delta_y(\bkf)$, they are given by
\begin{align}
&u^{\rm B}(r) = \sqrt{\frac{|\Delta_x(\bkf)|}{2\vf}} \re^{-\frac{|\Delta_x(\bkf)|}{\vf}|r|}, \cr
&v^{\rm B}(r) = -\ri \sqrt{\frac{|\Delta_x(\bkf)|}{2\vf}} \re^{-\frac{|\Delta_x(\bkf)|}{\vf}|r|}.
\label{eqn:uv-pxy}
\end{align}
Substituting Eq. (\ref{eqn:uv-pxy}) into Eq. (\ref{eqn:odd-f-pxy}), we obtain
\begin{align}
&\frac{1}{\ri\om-\Delta_y(\bkf)} u^{\rm B}(r) [v^{\rm B}(r)]^* \cr
&= \frac{1}{\vf} \frac{|\Delta_x(\bkf)|}{\om^2+\Delta_y^2(\bkf)} \left[ \om - \ri \Delta_y(\bkf) \right]
                                                                 \re^{-\frac{2|\Delta_x(\bkf)|}{\vf}|r|}. \cr
\end{align}
Here, the $\om$ term in the numerator represents the odd-frequency amplitude,
while the $-\ri\Delta_y(\bkf)$ term is for the even-frequency amplitude.
When we add the contribution from the continuum states, the pair amplitudes are expressed as
\cite{Matsumoto-1999}
\begin{align}
&F_{\bkf\uparrow\downarrow}^{\rm odd}(x,\ri\om)
= \frac{1}{2\vf} \frac{\om|\Delta_x(\bkf)|}{\om^2+\Delta_y^2(\bkf)} \re^{-2qx}, \cr
&F_{\bkf\uparrow\downarrow}^{\rm even}(x,\ri\om)
= \frac{1}{2\vf} \frac{-\Delta_x(\bkf)}{\Omega} \left( 1 - \re^{-2qx} \right) \cr
&~~~~~~
+ \frac{1}{2\vf} \frac{-\ri\Delta_y(\bkf)}{\Omega} \left[ 1 + \frac{\Delta_x^2(\bkf)}{\om^2+\Delta_y^2(\bkf)} \re^{-2qx} \right], \cr
\label{eqn:F-pxy}
\end{align}
where
\begin{align}
&\Omega = \sqrt{ \om^2+\Delta_x^2(\bkf)+\Delta_y^2(\bkf) }, \cr
&q = \Omega / |\vfx|.
\end{align}
It is clear that the odd-frequency pair amplitude exists also in the chiral $p_x+\ri p_y$-wave state.
As in the $p_x$-wave case, the odd-frequency pair amplitude decreases exponentially with $x$.
The first and second terms in $F_{\bkf\uparrow\downarrow}^{\rm even}(x,\ri\om)$ represent
the $p_x$- and $p_y$-wave components, respectively.
We can see that the former vanishes at $x=0$, while the latter is enhanced.
It is consistent with the spatial dependence of the order parameters shown in Fig. \ref{fig:gap-pxy}(a).

\subsection{Symmetry of odd-frequency order parameter}
\label{subsec:symmetry}

In this subsection, we discuss the symmetry of the emergent odd-frequency order parameter.
As in Eqs. (\ref{eqn:pair}) and (\ref{eqn:odd-f-pxy}), the odd-frequency pair amplitude is generated by the bound state.
In the presence of a finite interaction for the odd-frequency component,
the pair amplitude is stabilized as a superconducting order parameter.
Since the bound state is formed by pair-breaking perturbations such as surface and domain wall,
we can understand that the odd-frequency order parameter is brought about by the pair-breaking effects.
In case of the surface and domain wall, the spin-triplet nature of the $p$-wave Cooper pair holds
when the quasi-particle is scattered by the surface or when it passes through the domain wall.
This means that the induced order parameter has the same spin-triplet property.
Concerning the symmetry of the orbital of the Cooper pair,
we saw in the previous section that the $s$-wave and $d_{xy}$-wave order parameters were stabilized
near the surface and domain wall, respectively.
This means that the orbital symmetry of the odd-frequency order parameter
depends on the geometry of the boundary relative to the symmetry of the unconventional order parameter.
We discuss this point below.

\begin{figure}[t]
\begin{center}
\includegraphics[width=3cm,clip]{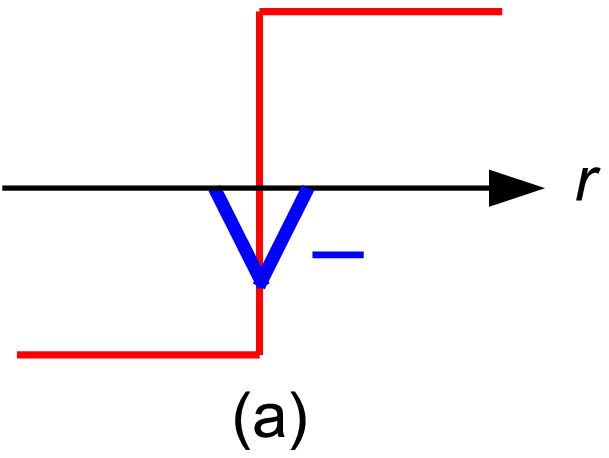}
\hspace{5mm}
\includegraphics[width=3cm,clip]{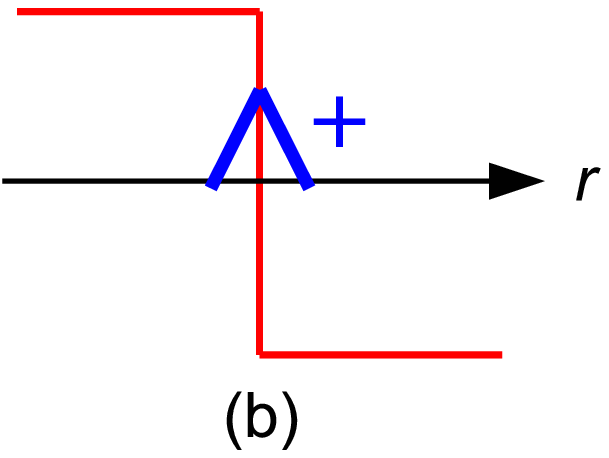}
\end{center}
\caption{
(Color online)
Order parameter [thin (red) lines] along the quasi-classical trajectory.
The solid (blue) lines represent the odd-frequency order parameter near the boundary.
The $\pm$ sign represents the sign of the order parameter.
(a) The case where the sign of the order parameter changes from negative to positive.
(b) From positive to negative case.
}
\label{fig:induced-order-parameter}
\end{figure}

We first summarize the result in the previous subsections.
When the sign of the order parameter changes from negative to positive along the quasi-classical trajectory,
a positive pair amplitude is induced.
For an attractive interaction,
the sign of the odd-frequency order parameter becomes negative as shown in Fig. \ref{fig:induced-order-parameter}(a).
In the opposite case, where the order parameter changes its sign from positive to negative,
the sign of the order parameter becomes positive [see Fig. \ref{fig:induced-order-parameter}(b)].
This is the key to understand the orbital symmetry of the odd-frequency order parameter.

\begin{figure}[t]
\begin{center}
\includegraphics[width=1.9cm,clip]{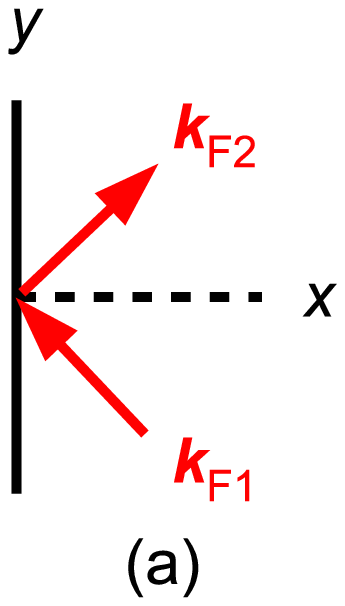}
\hspace{5mm}
\includegraphics[width=3.5cm,clip]{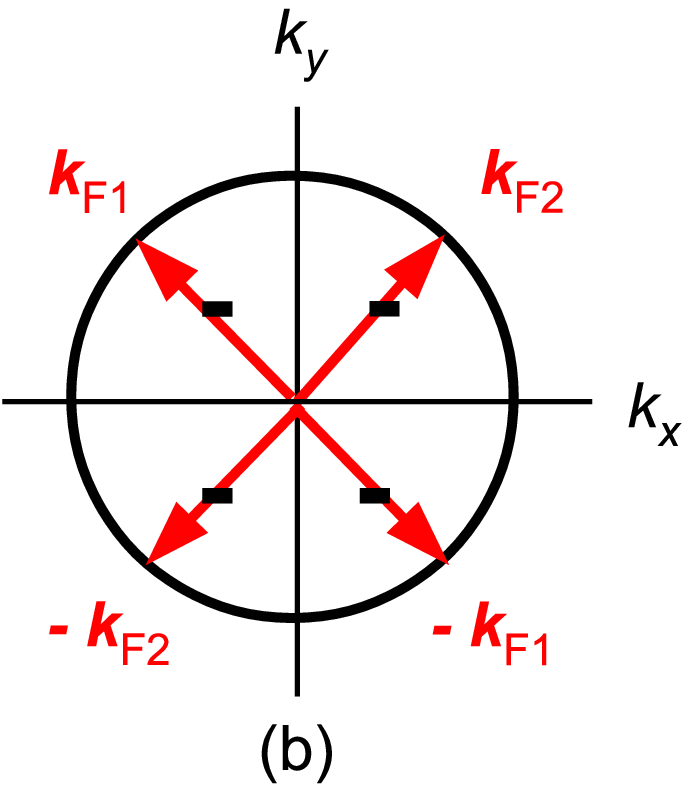} \\
\includegraphics[width=3cm,clip]{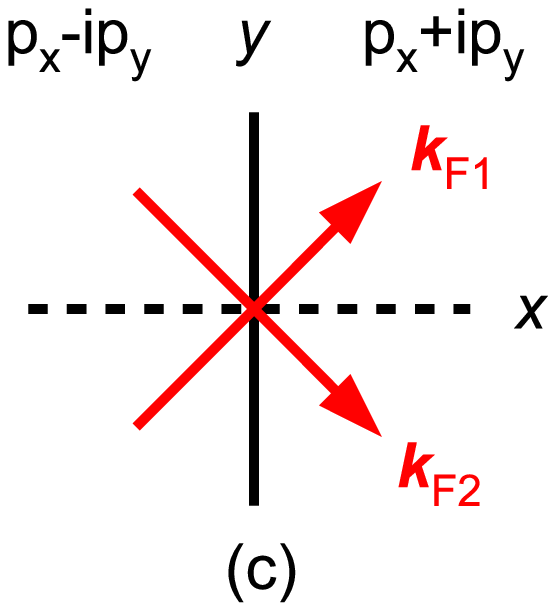}
\hspace{5mm}
\includegraphics[width=3.5cm,clip]{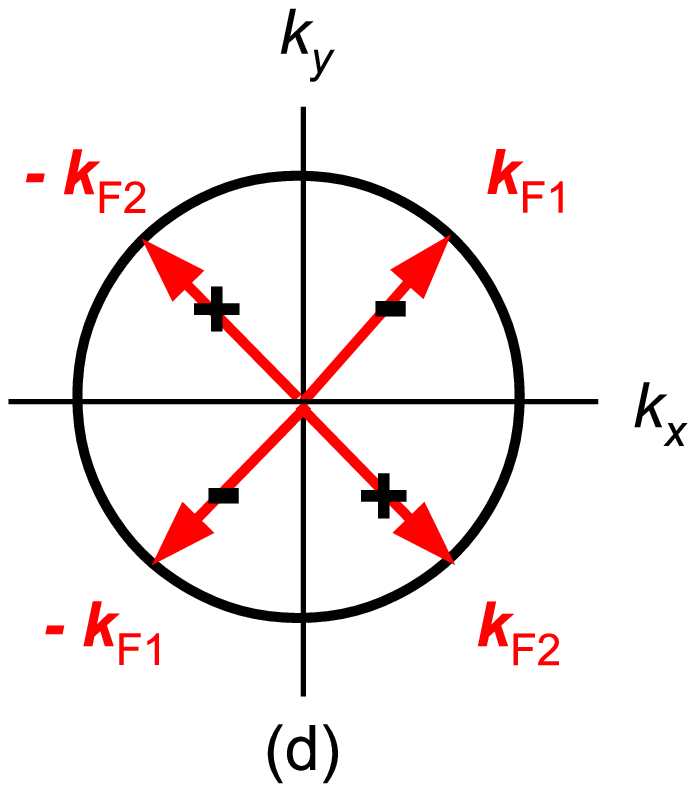}
\end{center}
\caption{
(Color online)
(a) Incident and reflected quasi-particles at the surface.
(b) Orbital symmetry ($s$-wave) of the odd-frequency order parameter.
(c) Incident and reflected quasi-particles for the domain wall.
(d) Orbital symmetry ($d_{xy}$-wave) of the odd-frequency order parameter.
}
\label{fig:scattering}
\end{figure}

Let us begin with the $p_x$-wave superconductor.
At the surface, there are two types of scatterings.
One is shown in Fig. \ref{fig:scattering}(a), while in the other one the directions of the wave vectors are reversed.
In both cases, the sign of the order parameter changes from negative to positive as shown in Fig. \ref{fig:induced-order-parameter}(a).
Therefore, the odd-frequency order parameter has an $s$-wave symmetry with a negative sign as shown in Fig. \ref{fig:scattering}(b).
This is consistent with the results in Fig. \ref{fig:gap-px}.
This property holds also in the $p_x+\ri p_y$-wave case,
since only the $p_x$-wave component carries the sign-change of the order parameter.

In case of the $p_x-\ri p_y|p_x+\ri p_y$ domain wall, the quasi-particle passes through the domain wall.
The sign-change of the order parameter occurs in the $p_y$-wave component in this case.
For $\bkf_1$ shown in Fig. \ref{fig:scattering}(c), the sign changes from negative to positive,
while it is opposite for $\bkf_2$.
The former and the latter cases correspond to the situation
shown in Figs. \ref{fig:induced-order-parameter}(a) and \ref{fig:induced-order-parameter}(b), respectively.
This leads to a $d_{xy}$-wave symmetry of the odd-frequency order parameter
with a negative sign of the global phase factor ($\pi$ phase shift relative to the sign-change even-frequency component)
as shown in Fig. \ref{fig:scattering}(d).
This is also consistent with the results shown in Fig. \ref{fig:gap-pxy-domain}.
Since the sign-change $p_y$-wave component is pure imaginary for the $p_x-\ri p_y|p_x+\ri p_y$ domain wall,
we note that the odd-frequency $d_{xy}$-wave order parameter becomes also pure imaginary.

\section{Summary and Discussions}

In this paper, we investigated the emergent odd-frequency superconducting order parameter
near boundaries in unconventional superconductors.
We focus on the two-dimensional spin-triplet even-frequency $p_x$- and $p_x+\ri p_y$-wave superconducting states
of the $\bd \parallel z$ type, where the latter is the candidate of the paring state for \Sr~superconductor.

In the $p_x$-wave superconductors with a specular surface along the $y$-direction,
it is known that the order parameter is suppressed and the mid-gap bound state appears near the surface.
In the previous studies, it was suggested that the odd-frequency spin-triplet $s$-wave pair amplitude is induced near the surface.
\cite{Tanaka-2012}
In our study, we took the idea one step further and investigated the odd-frequency order parameter
by introducing a finite interaction in the odd-frequency pairing channel.
In particular, we considered the electron-phonon interaction,
since it leads to a finite attractive interaction in the spin-triplet $s$-wave channel.
We extended the quasi-classical theory to include the frequency-dependence of the order parameter
and confirmed that the conventional relation between the particle and hole components of the order parameter,
i.e. Eq. (\ref{eqn:condition}), must be used when we consider the
induced odd-frequency order parameter with the primary even-frequency order parameter.
Solving the extended Eilenberger equation self-consistently,
we determined both the frequency and spatial dependences of the order parameter
and found that the odd-frequency spin-triplet $s$-wave superconducting order parameter appears near the surface
(see Figs. \ref{fig:gap-px} and \ref{fig:gap-pxy}).

To understand the emergence of the odd-frequency order parameter,
we expressed the pair amplitude with the quasi-particle wavefunctions as given by Eq. (\ref{eqn:pair}).
It clearly indicates that the mid-gap bound state generates the odd-frequency pair amplitude.
In case of the chiral $p_x+\ri p_y$-wave superconductor, the bound state energy is located at $E=\Delta_y(\bkf)$ for each ${\kf}_y$
reflecting the broken time-reversal symmetry.
Although they are not the mid-gap state, the surface bound state generates the odd-frequency pair amplitude.
Under a finite interaction, the pair amplitude is stabilized and the odd-frequency order parameter appears.

When we compare the magnitude of the order parameters,
we notice that the odd-frequency component is quite small even under a strong attractive interaction in the odd-frequency channel.
The main reason for this is that the odd-frequency order parameter is hard to be stabilized in the bulk
owing to the limiting frequency region for the attractive interaction.
\cite{Kusunose-2011-1}
This property remains also in the non-uniform case of the induced odd-frequency order parameter near boundaries.
The other reason is that the frequency dependence in the bulk $p$-wave order parameter was not taken into account
under assuming the weak coupling theory.
In this case, the $p$-wave component is overestimated and dominates the minority $s$-wave odd-frequency component.
Although the magnitude of the odd-frequency order parameter is small,
we elucidated how it appears near boundaries in unconventional superconductors.
It appears as long as the coupling is finite irrespective of its sign (attractive or repulsive)
as shown in Figs. \ref{fig:gap-temp-px} and \ref{fig:gap-repulsive}.
This point can be explained qualitatively by the third-order term in the free energy given in Eq. (\ref{eqn:free-energy}).
When the odd-frequency order parameter is finite, physical quantities show different behavior.
Although the effect is expected to be weak in case of a small odd-frequency order parameter,
it is important to study this point in a general case.

In this paper, we focussed on surface and domain wall in unconventional superconductors.
The idea of the emergent odd-frequency order parameter can be adopted
also to interfaces between normal metal and unconventional superconductors
or between ferromagnet and superconductors and so on, since the odd-frequency pair amplitude is induced in these cases.
\cite{Bergeret-2005,Tanaka-2012}
In particular, in the former case, a larger magnitude of the odd-frequency order parameter is expected in the normal metal side,
since the unconventional (even-frequency) order parameter vanishes in the normal metal
and there is no competition between the even- and odd-frequency ones.
In this case, we can expect that the effect of the odd-frequency order parameter is seen in the normal metal side.
These points will be examined in the future work.

\section*{Acknowledgment}

This work is supported by a Grant-in-Aid for Scientific Research C (No. 23540414)
from the Japan Society for the Promotion of Science.
One of the authors (H.K.) is supported by a Grant-in-Aid for Scientific Research on Innovative Areas “Heavy Electrons” (No. 20102008) of the Ministry of Education, Culture, Sports, Science, and Technology (MEXT), Japan.

\appendix
\setcounter{equation}{0}

\section{Derivation of Andreev Equation}
\label{appendix:Andreev}

In case of the $p_x$-wave of the $\bd$-vector parallel to the $z$-axis type,
the field operators satisfy the following equation of motion:
\begin{align}
&\frac{\partial}{\partial \tau}
\left(
  \begin{matrix}
    \psi_\uparrow(\br,\tau) \cr
    \psi_\downarrow^\dagger(\br,\tau)
  \end{matrix}
\right)
= - \int \rd \br' \cr
&\times
\left(
  \begin{matrix}
    \delta(\br-\br') \varepsilon(-\ri\nabla') & \Delta(\br,\br') \cr
    - \Delta^*(\br,\br') & - \delta(\br-\br') \varepsilon(-\ri\nabla')
  \end{matrix}
\right)
\left(
  \begin{matrix}
    \psi_\uparrow(\br',\tau) \cr
    \psi_\downarrow^\dagger(\br',\tau)
  \end{matrix}
\right). \cr
\label{eqn:BdG}
\end{align}
Here, $\psi_\uparrow(\br,\tau)$ and $\psi_\downarrow^\dagger(\br,\tau)$ are field operators in the Heisenberg representation defined by
\begin{align}
\left(
  \begin{matrix}
    \psi_\uparrow(\br,\tau) \cr
    \psi_\downarrow^\dagger(\br,\tau)
  \end{matrix}
\right)
=
\re^{\mathcal{H}\tau}
\left(
  \begin{matrix}
    \psi_\uparrow(\br) \cr
    \psi_\downarrow^\dagger(\br)
  \end{matrix}
\right)
\re^{-\mathcal{H}\tau},
\end{align}
where $\mathcal{H}$ is the Hamiltonian.
$\Delta(\br,\br')$ is the position dependent order parameter.
$\varepsilon(-\ri\nabla')=(-\ri\nabla')^2/(2m) - E_{\rm F}$ is the operator for the kinetic energy
measured from the Fermi energy $E_{\rm F}$.

To derive the Andreev equation, we rewrite the field operators as
\begin{align}
\left(
  \begin{matrix}
    \psi_\uparrow(\br) \cr
    \psi_\downarrow^\dagger(\br)
  \end{matrix}
\right)
= \sum_\bkf
\left(
  \begin{matrix}
    \psi_{\bkf\uparrow}(\br) \cr
    \psi_{-\bkf\downarrow}^\dagger(\br)
  \end{matrix}
\right) \re^{\ri \bkf\cdot\br}.
\label{eqn:psi-k}
\end{align}
The field operators $\psi_\uparrow(\br)$ and $\psi_\downarrow^\dagger(\br)$
contain the rapid oscillation term $\re^{\ri \bkf\cdot\br}$,
while the field operators $\psi_{\bkf\uparrow}(\br)$ and $\psi_{-\bkf\downarrow}^\dagger(\br)$ describe
slowly varying component of the order of the superconducting coherence length.
We substitute Eq. (\ref{eqn:psi-k}) into Eq. (\ref{eqn:BdG}) and apply the quasi-classical approximation,
\cite{Andreev,Eilenberger}
where the relative coordinate $\bx=\br-\br'$ and the center of mass coordinate $\bR=(\br+\br')/2$ are introduced.
The two coordinates, $\br$ and $\br'$, are then expressed as $\br=\bR+\bx/2$ and $\br'=\bR-\bx/2$.
The order parameter is written by $\bx$ and $\bR$ as
\begin{align}
\Delta(\br,\br') \rightarrow \Delta(\bx,\bR).
\end{align}
The order parameter and the slowly varying field operator can be expanded as
\cite{Bruder}
\begin{align}
&\Delta(\bx,\bR) = \Delta(\bx,\br-\frac{\bx}{2}) \simeq \Delta(\bx,\br) - \frac{\bx}{2}\cdot\nabla\Delta(\bx,\br) + \cdots, \cr
&\psi_\bkf(\br') = \psi_\bkf(\br-\bx) \simeq \psi_\bkf(\br) - \bx\cdot\nabla\psi_\bkf(\br) + \cdots.
\label{eqn:expand}
\end{align}
For a fixed $\bkf$, we substitute Eq. (\ref{eqn:expand}) into Eq. (\ref{eqn:BdG}) and obtain
\cite{Bruder}
\begin{align}
&\int \rd \br' \Delta(\br,\br') \psi_{\downarrow}^\dagger(\br')
\rightarrow \int \rd \br' \Delta(\bx,\bR) \psi_{\bkf\downarrow}^\dagger(\br') \re^{\ri\bkf\cdot\br'} \cr
&= \re^{\ri\bkf\cdot\br} \int \rd \br' \Delta(\bx,\br-\frac{\bx}{2}) \psi_{\bkf\downarrow}^\dagger(\br-\bx)
                                       \re^{-\ri\bkf\cdot(\br-\br')} \cr
&\simeq \re^{\ri\bkf\cdot\br} \int \rd \bx
\left[ \Delta(\bx,\br) - \frac{\bx}{2}\cdot\nabla\Delta(\bx,\br) \right] \cr
&~~~~~~~~~~~~~~~~\times
\left[ \psi_\bkf(\br) - \bx\cdot\nabla\psi_\bkf(\br) \right] \re^{-\ri\bkf\cdot\bx} \cr
&\simeq \re^{\ri\bkf\cdot\br} \int \rd \bx \Delta(\bx,\br) \psi_\bkf(\br) \re^{-\ri\bkf\cdot\bx} \cr
&= \re^{\ri\bkf\cdot\br} \Delta(\bkf,\br) \psi_\bkf(\br),
\end{align}
where we introduced
\begin{align}
\Delta(\bkf,\br) = \int \rd \bx \Delta(\bx,\br) \re^{-\ri\bkf\cdot\bx}
\end{align}
as the Fourier transformed order parameter with respect to the relative coordinate of the Cooper pair.
In $\Delta(\bkf,\br)$, the $\bkf$ dependence represents the orbital symmetry of the Cooper pair,
while the slowly varying $\br$ dependence represents the spatial dependence of the order parameter
in terms of the center of mass coordinate.
Then, we obtain the following equation of motion of the field operators:
\cite{Bruder}
\begin{align}
&\frac{\partial}{\partial \tau}
\left(
  \begin{matrix}
    \psi_{\bkf\uparrow}(\br,\tau) \cr
    \psi_{-\bkf\downarrow}^\dagger(\br,\tau)
  \end{matrix}
\right) \cr
&= -
\left(
  \begin{matrix}
    - \ri\bvf\cdot\nabla & \Delta(\bkf,\br) \cr
    \Delta(\bkf,\br) & \ri\bvf\cdot\nabla
  \end{matrix}
\right)
\left(
  \begin{matrix}
    \psi_{\bkf\uparrow}(\br,\tau) \cr
    \psi_{-\bkf\downarrow}^\dagger(\br,\tau)
  \end{matrix}
\right),
\label{eqn:Andreev-0}
\end{align}
where we used $-\Delta^*(-\bkf,\br) = \Delta(\bkf,\br)$ for the $p_x$-wave with a real value.
$\bvf=\bkf/m$ represents the Fermi velocity.
To find energy eigenstates, we put the field operators as
\begin{align}
\left(
  \begin{matrix}
    \psi_{\bkf\uparrow}(\br) \cr
    \psi_{-\bkf\downarrow}^\dagger(\br)
  \end{matrix}
\right)
=
\left(
  \begin{matrix}
    u_E(\br) \cr
    v_E(\br)
  \end{matrix}
\right) \gamma_E.
\label{eqn:psi-uv}
\end{align}
Here, $\gamma_E$ is a fermion operator for an energy eigenvalue $E$.
We assume that the Hamiltonian has a diagonalized form as $E\gamma_E^\dagger \gamma_E$.
Substituting Eq. (\ref{eqn:psi-uv}) into Eq. (\ref{eqn:Andreev-0}), we obtain the following Andreev equation:
\begin{align}
\left(
  \begin{matrix}
    - \ri\bvf\cdot\nabla & \Delta(\bkf,\br) \cr
    \Delta(\bkf,\br) & \ri\bvf\cdot\nabla
  \end{matrix}
\right)
\left(
  \begin{matrix}
    u_E(\br) \cr
    v_E(\br)
  \end{matrix}
\right)
= E
\left(
  \begin{matrix}
    u_E(\br) \cr
    v_E(\br)
  \end{matrix}
\right).
\label{eqn:Andreev}
\end{align}

\section{Boundary Condition of Andreev Equation}
\label{appendix:boundary-condition}

\begin{figure}[t]
\begin{center}
\includegraphics[width=3.5cm,clip]{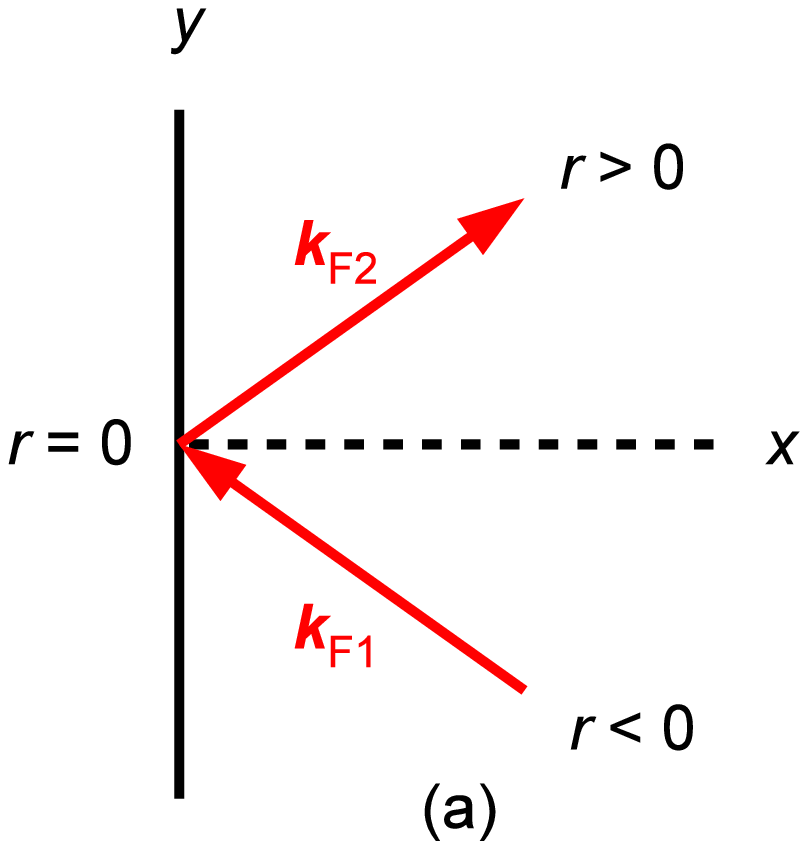}
\hspace{5mm}
\includegraphics[width=4cm,clip]{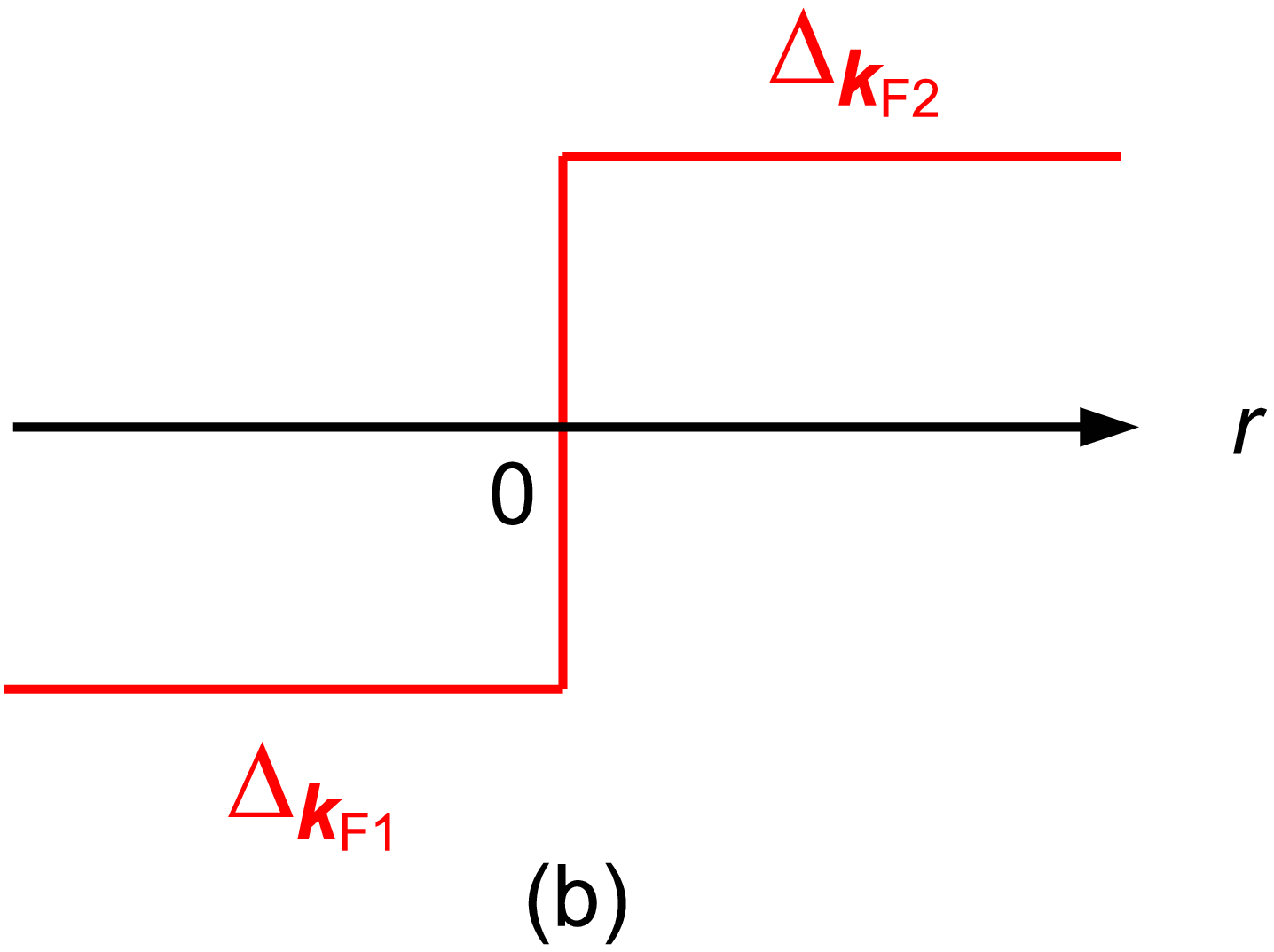}
\end{center}
\caption{
(Color online)
(a) Quasi-classical trajectory.
(b) Spatial dependence of the order parameter along the quasi-classical trajectory.
}
\label{fig:boundary}
\end{figure}

In this appendix, we discuss the boundary condition of the Andreev equation at the surface.
The quasi-classical trajectory is shown in Fig. \ref{fig:boundary}(a).
We take the coordinate $r$ along the quasi-classical trajectory.
The incident quasi-particle with $\bkf_1$ wavevector comes from $r=-\infty$ and is reflected at the surface at $r=0$.
The reflected quasi-particle travels with $\bkf_2$ wavevector toward $r=\infty$.
The order parameter for the incident quasi-particle is $\Delta_{\bkf_1}$,
while it is $\Delta_{\bkf_2}$ for the reflected quasi-particle.
For an uniform order parameter, the spatial dependence of the order parameter along the quasi-classical trajectory
suddenly changes its sign at the surface as shown in Fig. \ref{fig:boundary}(b).
We solve the Andreev equation (\ref{eqn:Andreev}) in the two regions separately and obtain the solutions analytically.

Let us discuss the boundary condition of the wavefunction of the quasi-particle.
In the presence of the specular surface along the $y$-direction, the $y$ component of the wavevector ($\kfy$) is conserved.
The $y$ dependence of the wavefunction is then expressed by $\re^{\ri\kfy y}$.
On the other hand, the $\pm\kfx$ components are coupled by the surface.
For a fixed $\kfy$, the field operator is written as
\begin{align}
\left(
  \begin{matrix}
    \psi_{\uparrow}(\br) \cr
    \psi_{\downarrow}^\dagger(\br)
  \end{matrix}
\right)
=
\left(
  \begin{matrix}
    \psi_{\bkf_1\uparrow}(x) \re^{-\ri\kfx x} + \psi_{\bkf_2\uparrow}(x) \re^{\ri\kfx x} \cr
    \psi_{-\bkf_1\downarrow}^\dagger(x) \re^{-\ri\kfx x} + \psi_{-\bkf_2\downarrow}^\dagger(x) \re^{\ri\kfx x} \cr
  \end{matrix}
\right) \re^{\ri \kfy y}.
\end{align}
Here, $\bkf_1=(-\kfx,\kfy)$ and $\bkf_2=(\kfx,\kfy)$.
The $y$ dependence is extracted as $\re^{\ri\kfy y}$.
The field operators $\psi_{\bkf\uparrow}(x)$ and $\psi_{-\bkf\downarrow}^\dagger(x)$ only have the $x$ dependence.
Since the field operator must vanish at the surface, we obtain
\begin{align}
&\psi_{\bkf_1\uparrow}(0) + \psi_{\bkf_2\uparrow}(0) = 0, \cr
&\psi_{-\bkf_1\downarrow}^\dagger(0) + \psi_{-\bkf_2\downarrow}^\dagger(0) = 0,
\label{eqn:condition-psi}
\end{align}
for each $\kfy$.
These lead to the following boundary condition for $u$ and $v$ at $r=0$:
\begin{align}
\left(
  \begin{matrix}
    u(0) \cr
    v(0)
  \end{matrix}
\right)_{\bkf_1}
= -
\left(
  \begin{matrix}
    u(0) \cr
    v(0)
  \end{matrix}
\right)_{\bkf_2}.
\label{eqn:condition-uv0}
\end{align}
Since the Andreev equation (\ref{eqn:Andreev}) is invariant under the change of the global phase of the wavefunctions,
the minus sign in Eq. (\ref{eqn:condition-uv0}) can be absorbed in the wavefunctions.
Therefore, we solve the Andreev equation in both $r<0$ and $r>0$ regions separately
and connect the solutions with the following conditions at $r=0$:
\begin{align}
\left(
  \begin{matrix}
    u(0) \cr
    v(0)
  \end{matrix}
\right)_{\bkf_1}
=
\left(
  \begin{matrix}
    u(0) \cr
    v(0)
  \end{matrix}
\right)_{\bkf_2}.
\label{eqn:condition-uv}
\end{align}
This indicates that it is sufficient to solve the Andreev equation (\ref{eqn:Andreev}) continuously
along the quasi-classical trajectory.
In the similar way, the quasi-classical Green's function is also solved continuously along the quasi-classical trajectory
as we mentioned in Eq. (\ref{eqn:boundary-G}).



\begin{thebibliography}{99}
\bibitem{note-L}
  We assumed that there are two boundaries at $r'=\pm L/2$ for the Andreev equation on the quasi-classical trajectory.
  To be precise, the size of the system $L$ depends on the direction of $\bkf$.
  Since the final result does not depend on the size of the system, we use the same $L$ for various $\bkf$.
\end{thebibliography}
\end{document}